\documentclass[pre,aps,twocolumn,floatfix,showpacs,superscriptaddress]{revtex4-1}
\usepackage{amssymb}
\usepackage{amsmath}
\usepackage{graphicx}
\usepackage{physics}
\usepackage[usenames]{color}
\begin{document}

\title{Manifestation of strange nonchaotic attractors in extended systems: A study through out-of-time-ordered correlators}

\author{P. Muruganandam}
\affiliation{Department of Physics, Bharathidasan University, Tiruchirappalli 620024, Tamil Nadu, India}

\author{M. Senthilvelan}
\affiliation{Department of Nonlinear Dynamics, Bharathidasan University, Tiruchirappalli 620024, Tamil Nadu, India}

\begin{abstract}
We study the spatial spread of out-of-time-ordered correlators (OTOCs) in coupled map lattices (CMLs) of quasiperiodically forced nonlinear maps. We use instantaneous speed (IS) and finite-time Lyapunov exponents (FTLEs) to investigate the role of strange non-chaotic attractors (SNAs) on the spatial spread of the OTOC. We find that these CMLs exhibit a characteristic on and off type of spread of the OTOC for SNA. Further, we provide a broad spectrum of the various dynamical regimes in a two-parameter phase diagram using IS and FTLEs. We substantiate our results by confirming the presence of SNA using established tools and measures, namely the distribution of finite-time Lyapunov exponents, phase sensitivity, spectrum of partial Fourier sums, and $0-1$ test.
\end{abstract}

\maketitle

\section{Introduction}

The delocalization of information over a physical system, after which the information is inaccessible to the local measurements, is called information scrambling \cite{Swingle2016, Joshi2020, Bergamasco2020, Sreeram2021, Sreeram2021a}. In an isolated quantum system, this spreading can be estimated by the entanglement entropy \cite{Luitz2017, Ghosh2019}, tripartite mutual information \cite{Hosur2016} and out-of-time-ordered correlators (OTOC) \cite{Swingle2016, Joshi2020, Luitz2017, Ghosh2019}.
For two local Hermitian operators ($\hat{P} $ and $\hat{R}$), the quantum OTOC is defined as \cite{Zhao2021}
\begin{align}
\expval{C(t)} = \expval{[\hat{P}(t),\hat{R}(0)]^{2}},
	    \label{def_OTOC}
\end{align}
where $\expval{\cdot}$ represents the average of the initial state and  $P(t)= e^{iHt}\hat P(0)e^{-iHt}$ is the time dependent operator in the Heisenberg picture  \cite{Zhao2021}. At no scrambling stage ($t=0)$, the operators that encode the initial mode of quantum information commute. As time goes on, the commutativity may get broken, expressing the growing non-locality and complexity of the encoded information, which can be interpreted as information scrambling in the considered system \cite{Sharma2021}.

In quantum systems, the rate at which the exponential growth of OTOC occurs is called the quantum Lyapunov exponent and has been considered as a quantum counterpart of classical chaos \cite{Xu2020}. The OTOC diagnoses the ballistic spread of local information in quantum systems \cite{Das2018}. The OTOC measure can be related to other chaos diagnostics such as Loschmidt echo and complexity \cite{Bhattacharyya2021, Bergamasco2019}. The exponential growth may occur due to the presence of unstable fixed points and not due to chaos \cite{Xu2020, PilatowskyCameo2020}. One may note that the systems which exhibit exponential growth of OTOC do not necessarily admit chaos when the potential includes a local maximum \cite{Hashimoto2020}. In recent years studies have been made on understanding OTOC from different perspectives. A comparison between the early growth of OTOCs in integrable and chaotic Ising chains have been made under two different boundary conditions and it is observed that both the integrable and chaotic systems have (i) \textit{very close} early growth under periodic boundary condition and (ii) they have \textit{close} early growth under open boundary condition \cite{Yan2019}. It has also been demonstrated that the classical-to-quantum correspondence in OTOC dynamics is being violated in certain nonchaotic quantum and classical systems \cite{Rozenbaum2020}. In quantum systems, the early-times growth resembles Lyapunov-like exponential with Planck's constant dependent rate. This result strongly contradicts their classical counterparts, which show slow early-time growth \cite{Rozenbaum2020}. 

It has also been found that both the short and long time behaviours of OTOC characterize the essential features of classical chaos \cite{GarciaMata2018}. The long time dynamics of OTOC helps to detect the transition between integrability and chaos in certain quantum maps and spin chains \cite{Fortes2019}. From the growth and long-time saturation of OTOC, the onset of chaos has been studied in the kicked Dicke model \cite{Sinha2021}. Further, a single qubit OTOC has been used to probe both the ground state quantum phase transitions (QPTs) and excited state QPTs in a many particle system \cite{Mumford2020}. The connection between OTOC and the dynamical QPTs has also been demonstrated through nuclear magnetic resonance quantum simulator experiments \cite{Nie2020}. The OTOCs have been experimentally detected in the presence of non-equilibrium phase transitions in traverse field Ising model \cite{Chen2020}. The OTOC also detects (i) the QPT between normal phase and superradiant phase and (ii) quantum critical points in the Rabi and Dicke models \cite{Sun2020}. Subsequently, based on the computation, the OTOC has been classified as microcanonical OTOC, thermal OTOC \cite{Romatschke2021} and fidelity OTOC \cite{PilatowskyCameo2020, LewisSwan2019}.

Very recently, OTOC also set its foot on portraying spatiotemporal chaos in classical extended systems. In the classical consideration, the commutator \ref{def_OTOC} is replaced by the Poisson bracket, $\{x_i(t),p_i(0)\}$, which approximates itself to $\frac{\delta x_i(t)}{\delta x_i(0)}$. Hence, Eq. \ref{def_OTOC} is approximated to $\expval{\abs{\frac{\delta x_i(t)}{\delta x_i(0)}}^2}$ and the $\expval{\cdot}$ refers to the thermal average \cite{Chatterjee2020}. Using the spatiotemporal heat maps of the OTOC the authors of Ref. \cite{Chatterjee2020} have demonstrated the existence of the three different dynamical regimes, namely (i) sustained chaos, (ii) nonchaotic regimes and (iii) transient chaos in the driven dissipative Duffing chain. In the sustained chaos regime, the OTOC shows exponential growth whereas in the non-chaotic regime it decays and in the transient region the OTOC has been found to exhibit a transition from sustained chaos to non-chaos \cite{Chatterjee2020}. Subsequently, the spatiotemporal spread of the OTOC in certain one-dimensional power-law models consisting of $N$ particles which are confined by an external harmonic potential has been investigated in depth in the linear regime (sufficiently at low temperatures and short times) \cite{Kiran2020}. The authors have computed the ground-state dispersion relation in the absence of external harmonic potential and obtained some analytical results for the OTOC. Interestingly, the analytical results which they found completely agreed with the numerical results. Further, specific features like butterfly speed has also been studied through collective field theory \cite{Kiran2020}. As in \cite{Chatterjee2020}, the space time heat maps of OTOC are found prominent to differentiate the ultra-low, low and high temperature regimes that prevail in the discrete nonlinear Schr\"{o}dinger chain \cite{Chatterjee2021}. 

In the present study, we exploit these out-of-time-ordered correlators to investigate the signatures of strange nonchaotic attractors (SNAs) in spatially extended systems. SNAs are attractors, which have fractal geometric structures but exhibit nonchaotic dynamics~\cite{Grebogi1984}. They occur most robustly in quasiperiodically-driven nonlinear oscillators, forced by incommensurate frequencies in the golden ratio. SNAs are not rare and found in many theoretical models and experiments, such as electronic circuits, plasma, neuronal membrane, quantum systems, stellar systems, etc.~\cite{Feudel2006, Ditto1990, Zhou1992, Mandell1993, Ding1997, Bondeson1985, Lindner2015}. SNAs are different from chaotic attractors in the sense that the neighbouring trajectories do not diverge exponentially in time. Due to this absence of sensitivity to initial conditions, SNA offers a prospect for predictability of the system’s state even from inaccurate observations, without drastic divergence in finite time.

We use coupled map lattices of quasiperiodically forced nonlinear maps, namely logistic and cubic maps, which are suitable candidates for studying SNAs on a spatially extended scale.  Notably, these coupled map lattices find applications in several areas such as dynamics, turbulence, phase transitions, geophysics, optics, genetics, human information processes, etc., due to their role as a family of systems possessing universal behaviour with an explicit form of the local interactions. The CMLs of quasiperiodically forced nonlinear maps are found to reflect the SNA properties of the isolated maps. We inspect the role of SNAs on the spatial spread of the OTOC by examining the instantaneous speed (IS) of the spread and the finite-time Lyapunov exponents (FTLEs). We substantiate the presence of SNA in the CML by carefully evaluating the established measures. 

This paper is organised as follows. In Sec.~\ref{sec:cml:1}, we study the coupled map lattices of quasiperiodically forced logistic maps with isolated maps exhibiting SNAs. First, we analyse the spread of the OTOC in a CML forced logistic maps by measuring the growth or decay of the initial perturbation given at one lattice, the IS of the spread, and the FTLEs  in different dynamical regimes of the CML. In the nonchaotic regime, we observe suppression of the spread of the initial perturbation after a finite number of iterations and the heat map of the OTOC shows balloon-like structure. In the chaotic regime, it shows ballistic spread with light-cone like structure.  Surprisingly,  we see an interesting type of \textit{on and off} spread of the OTOC in the SNA regime. We also study the behaviour of the IS and the FTLEs in all the three regimes and identify a two parameter phase diagram where different dynamical regions are clearly visible. Further, to confirm presence of SNA in the spatially extended systems, we carry out an extensive analysis using certain characteristic measures such as distribution of FTLEs, phase sensitivity, partial Fourier sums, and $0-1$ test.

In Sec.~\ref{sec:cml:2}, we consider the CML of quasiperiodically forced cubic maps and analyse the properties of OTOC. We analyse the nature of IS and FTLEs and provide a two parameter phase diagram. We further make a comprehensive study of the characteristic measures to confirm the presence of SNA in the CML. Finally, we conclude with a summary of the work in Sec.~\ref{sec:summary}.

\section{Coupled map lattices of quasiperiodically forced logistic maps}
\label{sec:cml:1}

We study the OTOC in coupled map lattices of quasiperiodic logistic maps of the following form
\begin{subequations}
\label{eq:cml:01}
\begin{align}
x_{n+1}^j = & \, (1 - \kappa)   f\left(x_n^{j}\right) + \frac{\kappa}{2} \left[ f\left(x_n^{j-1}\right)  +  f\left(x_n^{j+1}\right)  \right] , \\
\phi_{n+1} =& \,  \phi_n + \omega \; (\text{mod\ } 1), \;\;\; j = 0, \pm1, \pm2, \ldots M,
\end{align}
\end{subequations}
where $f(x) =  \alpha \left[ 1 +  \epsilon \cos 2 \pi \phi \right]  x  \left( 1 - x  \right)$ represents the quasiperiodically forced logistic map. 
We consider the above CML with $M=512$ ($2 M + 1 = 1025$ lattice points) and assume periodic boundary conditions or ring geometry. 

A single quasiperiodically forced logistic map is of the form~\cite{Prasad1997,Prasad1998,Prasad2001,Gopal2013} 
\begin{subequations}
\label{eq:cml:02}
\begin{align}
& x_{n+1} = \alpha \left[ 1 + \epsilon \cos 2 \pi \phi_n  \right] x_n \left( 1 - x_n \right) = f(x_n), \\
& \phi_{n+1} = \phi_n + \omega \; (\text{mod\ } 1), 
\end{align}
\end{subequations}%
where $\omega = \left( \sqrt{5} - 1\right) /2$ is the irrational driving frequency, and $\epsilon$ is the forcing amplitude. The  logistic map \ref{eq:cml:02} has been extensively studied for its SNA properties and it has been shown that the system \ref{eq:cml:02} exhibits different dynamical behaviours - namely, periodic, strange nonchaotic, and chaotic attractors~\cite{Prasad1997,Prasad2001}. For instance, by fixing the scaled forcing strength as $\epsilon' = 0.76$, where $\epsilon'  = \epsilon / (4/\alpha - 1)$, and the parameter $\alpha = 2.92$, $3.04$ and $3.10$, one can identify, respectively, quasiperiodic, strange non-chaotic, and chaotic types of dynamics.

\subsection{OTOC: Quasiperiodically forced logistic maps}

To measure OTOC,  we follow a similar approach described in Ref.~\cite{Chatterjee2020}. We let two sets of identical copies of the same system \ref{eq:cml:01} evolve with an infinitesimal difference ($\varepsilon$)  in the initial conditions  ($x^{j,I}_0$ and $x^{j,II}_0$)  from a chosen lattice point, the middle one $(j=0)$ for convenience. We then evaluate how this initial difference spread and grow spatiotemporally in the CML of forced logistic maps. We evaluate the spread by the OTOC, defined by~\cite{Chatterjee2020}
\begin{align}
D(j,n) = \frac{\left\vert x_n^{j,I} - x_n^{j,II} \right\vert}{\left\vert x_0^{0,I} - x_0^{0,II} \right\vert} =  \frac{\left\vert x_n^{j,I} - x_n^{j,II} \right\vert}{\lvert\varepsilon\rvert}. \label{eq:cml:otoc}
\end{align}

The quantity $D(j,n)$ essentially measures the ratio of the deviation between the two copies for the $j$-th lattice point at $n$-th iteration to the initial deviation at the middle lattice point, i.e. $\varepsilon$. Interestingly, $D(j,n)$  captures information on both the temporal growth (or decay) and spatial spread of the initial deviation. One may also define two quantities, namely the IS $v_b(n, D_{th})$ of the spatial spread and the FTLE, $\lambda_j(n)$, as 
\begin{align}
v_b(n, D_{th}) & = \frac{1}{n} \sum_{j=-M}^{M} \Theta\left[ D(j,n) - D_{th} \right], \label{eq:cml:vb}\\
\lambda_j(n) & = \frac{1}{n} \ln D(j,n), \label{eq:cml:le}
\end{align}
where $\Theta(\ldots)$ is the Heaviside step function. In the above, $v_b(n, D_{th})$ denotes the measure of number of lattice points per unit time ($n$) that have gained deviations greater or equal to $\varepsilon D_{th}$. The FTLE $\lambda_j(n)$ describes the local growth (or decay) of the deviation at a particular lattice point. It may be noted that, for systems exhibiting chaos, $\lambda_j(n)$ blows up for large $n$. So one has to use reorthonomalization procedure to compute the Lyapunov exponents~\cite{Wolf1985}. 

We first study the spatial spread of the OTOC in the CML of quasiperiodically driven logistic maps by estimating the OTOC $D(j,n)$, the instantaneous spread speed $v_b(n, D_{th})$ for different values of the threshold $D_{th}$, and the FTLEs $\lambda_j(n)$ at distinct lattice points. Our focus here is to see how the OTOC  characterises the three types of dynamics, namely quasiperiodic (nonchaotic), SNA and chaotic nature in the concerned system. Among the three, we concentrate more on SNA, which has not been studied so far.

In the limit $\varepsilon \to 0$, one can write the equation for the difference $\delta x_n^j = x_n^{j,I} - x_n^{j,II}$ as
\begin{align}
\delta x_{n+1}^j = & \, (1 - \kappa)   f'\left(x_n^{j}\right)  \delta x_n^{j}  + \frac{\kappa}{2} \left[ f'\left(x_n^{j-1}\right)   \delta x_n^{j-1}  
\right. \notag \\ & \, \left. 
+  f'\left(x_n^{j+1}\right)  \delta x_n^{j+1}  \right], j = 0, \pm 1, \pm 2, \ldots, M,
\label{eq:cml:lin01}
\end{align}
where $f'(x) = \alpha \left[ 1 +  \epsilon \cos 2 \pi \phi \right]   \left( 1 -2  x \right)$. We iterate the CML \ref{eq:cml:01} together with the linear equations \ref{eq:cml:lin01} starting from the initial conditions, say for example, $x_0^{j} = x_0^0$, $\delta x_0^{0} = \varepsilon$, and $\delta x_0^{j} = 0$ for $j \ne 0$. %
Alternatively, one can also iterate two sets of identical initial conditions with an infinitesimal difference in a chosen lattice point and use the difference between the two sets as $\delta x^j$ to estimate the OTOC. %
\begin{figure}[!ht]
\begin{center}
\includegraphics[width=\linewidth]{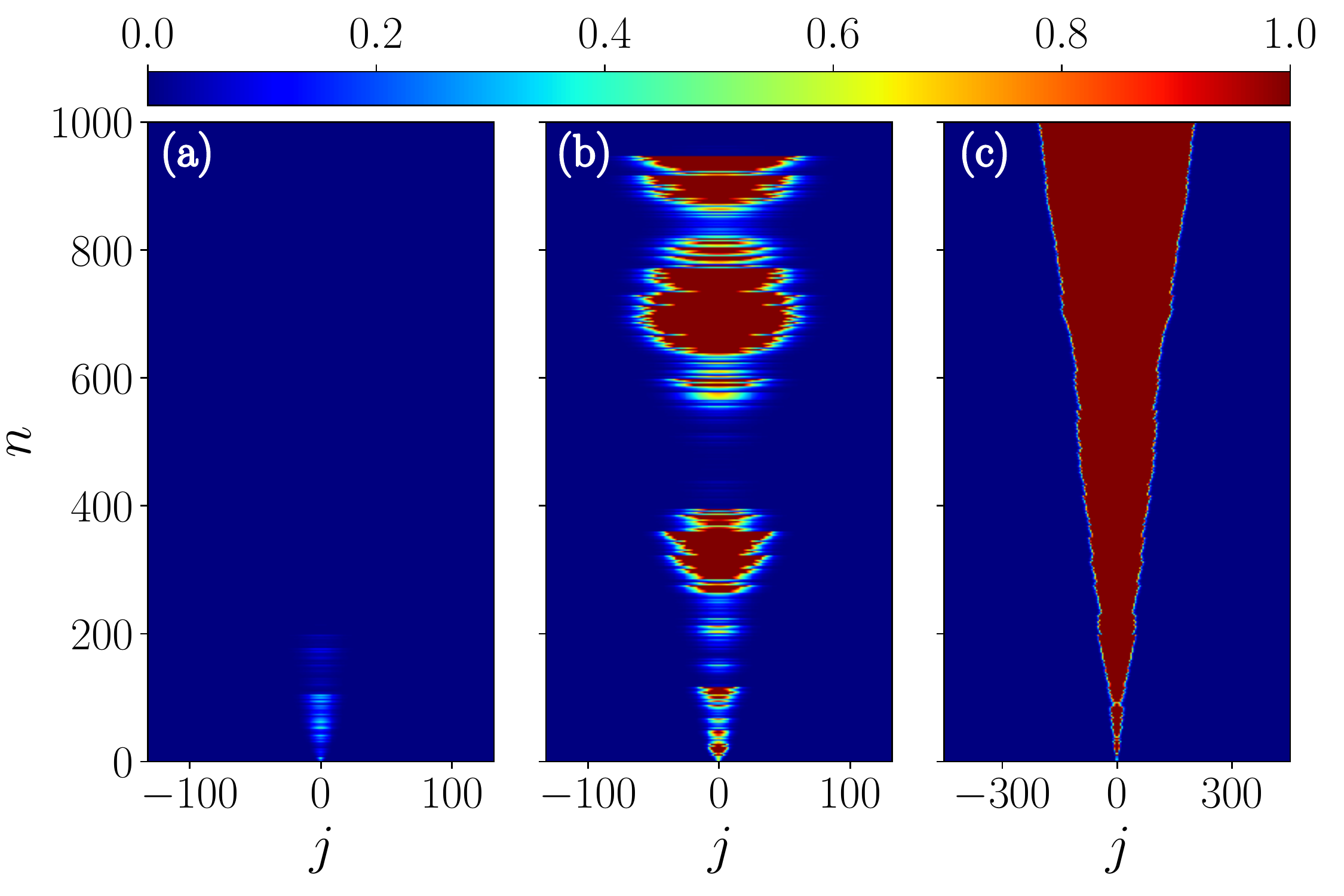}
\end{center}
\caption{Plot of the spatiotemporal spread of the perturbation in a CML of forced logistic maps in different dynamic regimes: (a) $\alpha = 2.92$ (non-chaotic), (b) $\alpha = 3.04$ (strange non-chaotic) and (c) $\alpha = 3.10$ (chaotic). The other parameters are $\epsilon' = 0.76$ and $\kappa = 0.5$ with $\epsilon = \epsilon' (4/\alpha - 1)$.}
\label{sna-logis}
\end{figure}%
In Fig.~\ref{sna-logis}, we depict a density contour (heat map) of the OTOC~\ref{eq:cml:otoc} in the $j - n$ plane. In the non-chaotic regime, the initial spread of the OTOC is suppressed after a finite number of iterations as shown in Fig.~\ref{sna-logis}(a). The initial growth and spread takes an exponential decay, and the spread forms a balloon shape. %
\begin{figure}[!ht]
\begin{center}
\includegraphics[width=\linewidth]{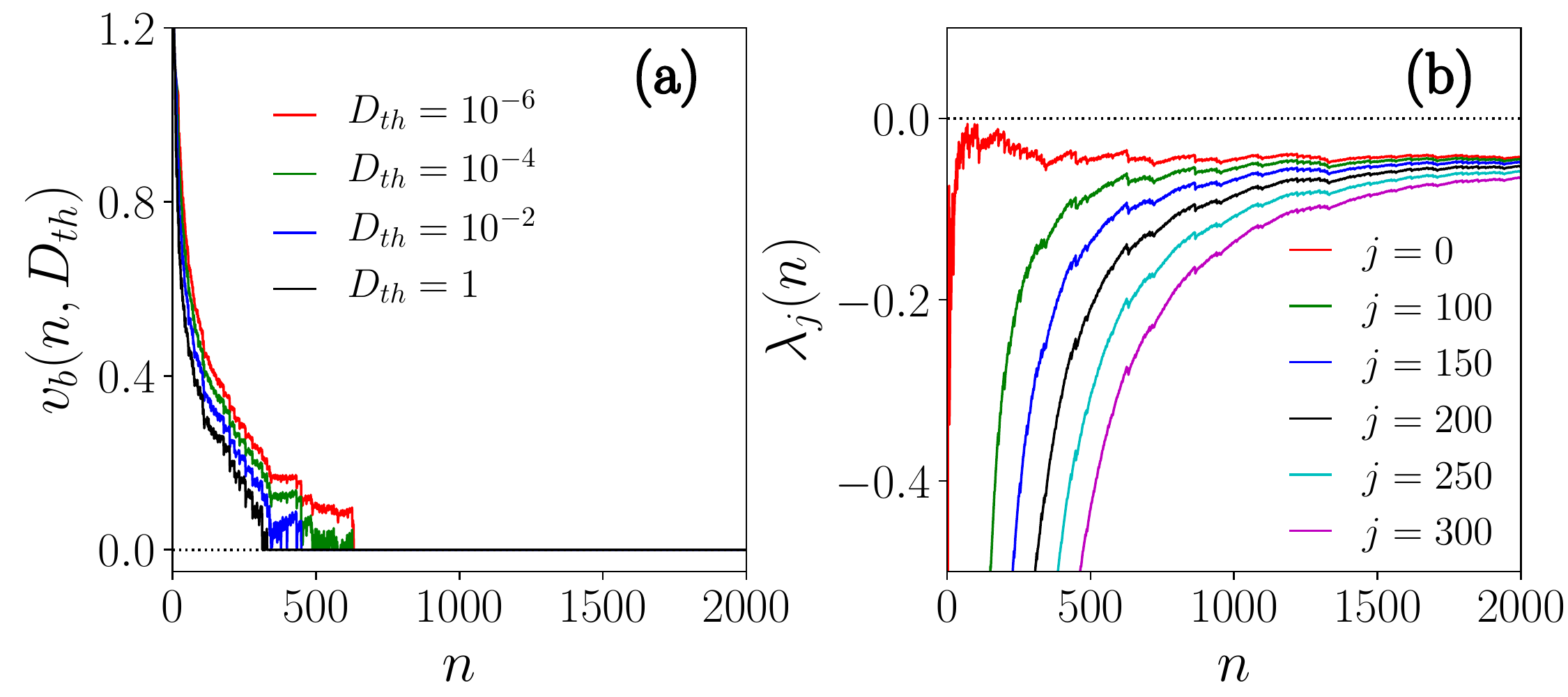}
\end{center}
\caption{CML (Logistic map):  (a) Instantaneous speed (IS) and (b) finite time Lyapunov exponent (FTLE) for different lattice points  in the case of non-chaotic attractor. The parameters are $\alpha = 2.92$, $\epsilon' = 0.76$, and $\kappa = 0.5$.}
\label{logis-lv-1}
\end{figure}%
The OTOC shows a ballistic spread for the chaotic case as seen in Fig.~\ref{sna-logis}(c).  In the chaotic regime, the heat map of $D(j,n)$ exhibits a light-cone like structure with sharp boundaries implying that the perturbation is propagating along the lattice. We observe that, in the nonchaotic and chaotic regimes, the CML of logistic maps exhibits the same kind of spread as reported for the chain of Duffing oscillators and non-interacting integrable Hamiltonian systems~\cite{Chatterjee2020}. On the other hand, in the SNA regime $(\alpha = 3.04)$, we observe an interesting dynamics of controlled spread and suppression of the OTOC as depicted in Fig.~\ref{sna-logis}(b).  In this case, the OTOC initially spreads to some lattices and then conceals completely. The spread starts again from another neighbouring lattice and grows for a while, and again gets suppressed completely and this process continues. This on and off spread can be observed indefinitely.  An intermittent like behaviour  between balloon (non-chaotic) and light-cone (chaotic) like spread can be observed in the OTOC in the SNA regime. Interestingly, this growth and death kind of spread can be seen commonly in a variety of systems that exhibit SNA. The heat map showing this type of controlled spread and suppression is being reported for the first time in the literature.

We compute the IS $(v_b)$ and the FTLE $(\lambda_j)$ using Eqs.~\ref{eq:cml:vb} and \ref{eq:cml:le}, respectively, for all three regimes. In Fig.~\ref{logis-lv-1}(a) we plot the IS for four different threshold values, i.e. $D_{th}  \in \{ 10^{-6}$, $10^{-4}$, $10^{-2}$, $1\}$, as a function of $n$.  The IS drops down to zero asymptotically for the nonchaotic case. We also show the variation of FTLE, calculated at different lattice points, as a function of $n$, in Fig.~\ref{logis-lv-1}(b), all of which converges to a fixed negative value, which confirms the nature of nonchaotic dynamics. Here the largest of the FTLE approaches to $\sim -10^{-1}$. 

\begin{figure}[!ht]
\begin{center}
\includegraphics[width=\linewidth]{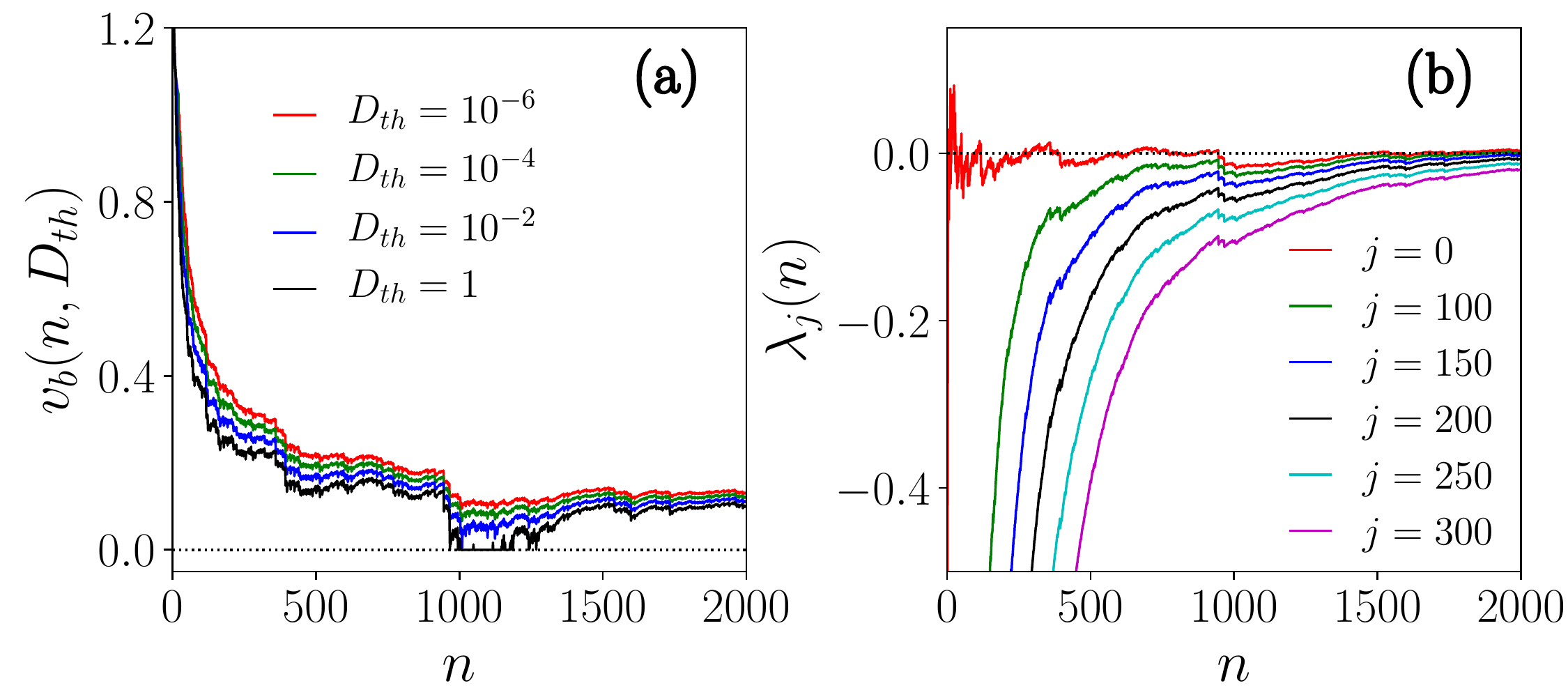}
\end{center}
\caption{CML (Logistic map): (a) Instantaneous speed (IS) and (b) finite time Lyapunov exponent (FTLE) for different lattice points  in the case of SNA. The parameters are $\alpha = 3.04$, $\epsilon'= 0.76$, and $\kappa = 0.5$.}
\label{logis-lv-2}
\end{figure}

Next, we consider the case in which the isolated logistic map \ref{eq:cml:02} exhibits SNA~\cite{Prasad2001}. We fix the parameters as  $\epsilon' = 0.76$ and $\alpha=3.04$,  The IS and FTLE are shown in Figs.~\ref{logis-lv-2}(a) and \ref{logis-lv-2}(b), respectively. %
Here we notice that the IS keeps on fluctuating in the vicinity of zero, and the FTLEs remain negative and the largest one is of the order of $-10^{-2}$. As we noted earlier, the isolated map  \ref{eq:cml:02} displays chaotic dynamics for the  parametric choice  $\epsilon' = 0.76$ and $\alpha=3.10$. %
\begin{figure}[!ht]
\begin{center}
\includegraphics[width=\linewidth]{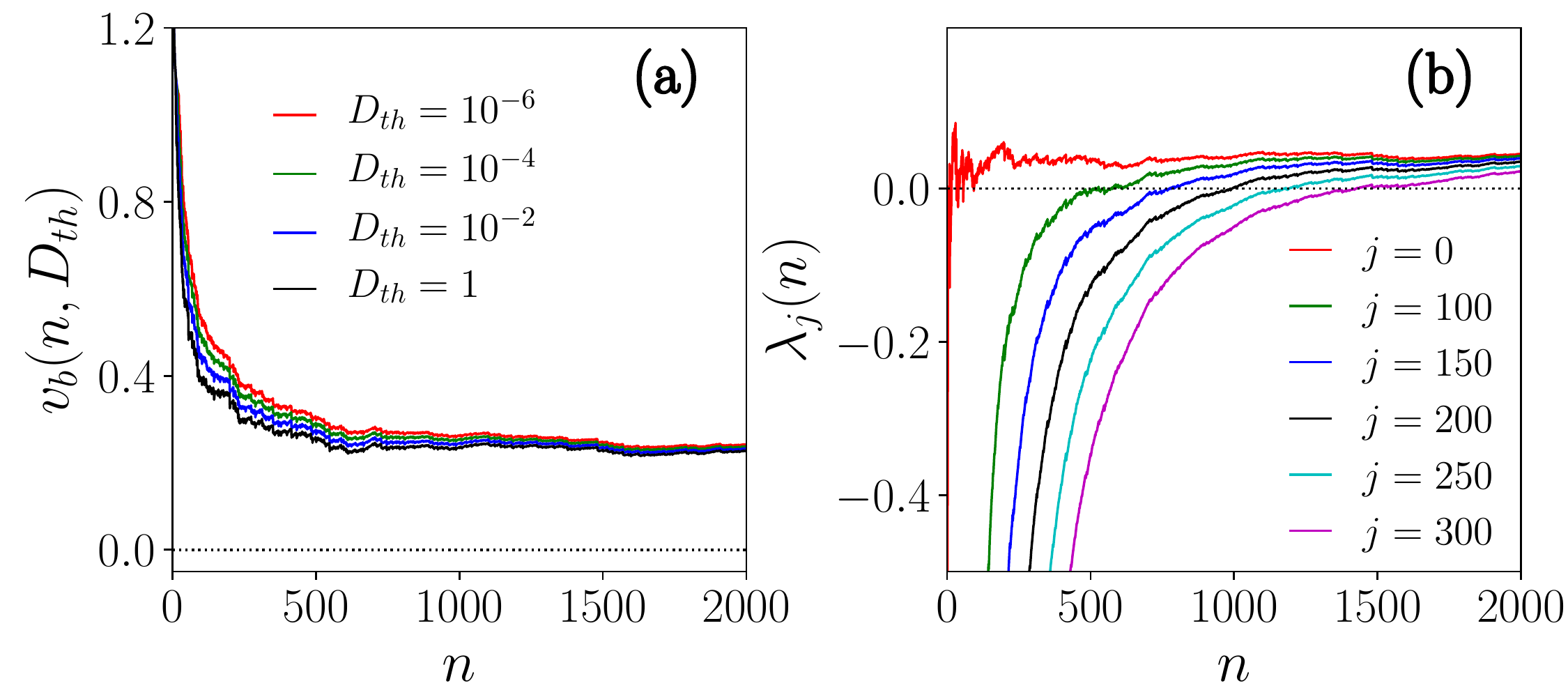}
\end{center}
\caption{CML (Logistic map): (a) Instantaneous speed (IS) and (b) finite time Lyapunov exponent (FTLE) for different lattice points in the case of chaotic attractor. The parameters are $\alpha = 3.10$, $\epsilon' = 0.76$, and $\kappa = 0.5$.}
\label{logis-lv-3}
\end{figure}%
In this case, we observe that the IS approaches a steady value, which is illustrated in Fig.~\ref{logis-lv-3}(a). In Fig.~\ref{logis-lv-3}(b). we plot the FTLE for various values of $j$. We observe that for large $n$, the FTLEs approach the value of $\lambda \sim 0.05$. In other words, the perturbation which we give in the middle lattice grows exponentially  with $n$. 

\begin{figure}[!ht]
\begin{center}
\includegraphics[width=\linewidth]{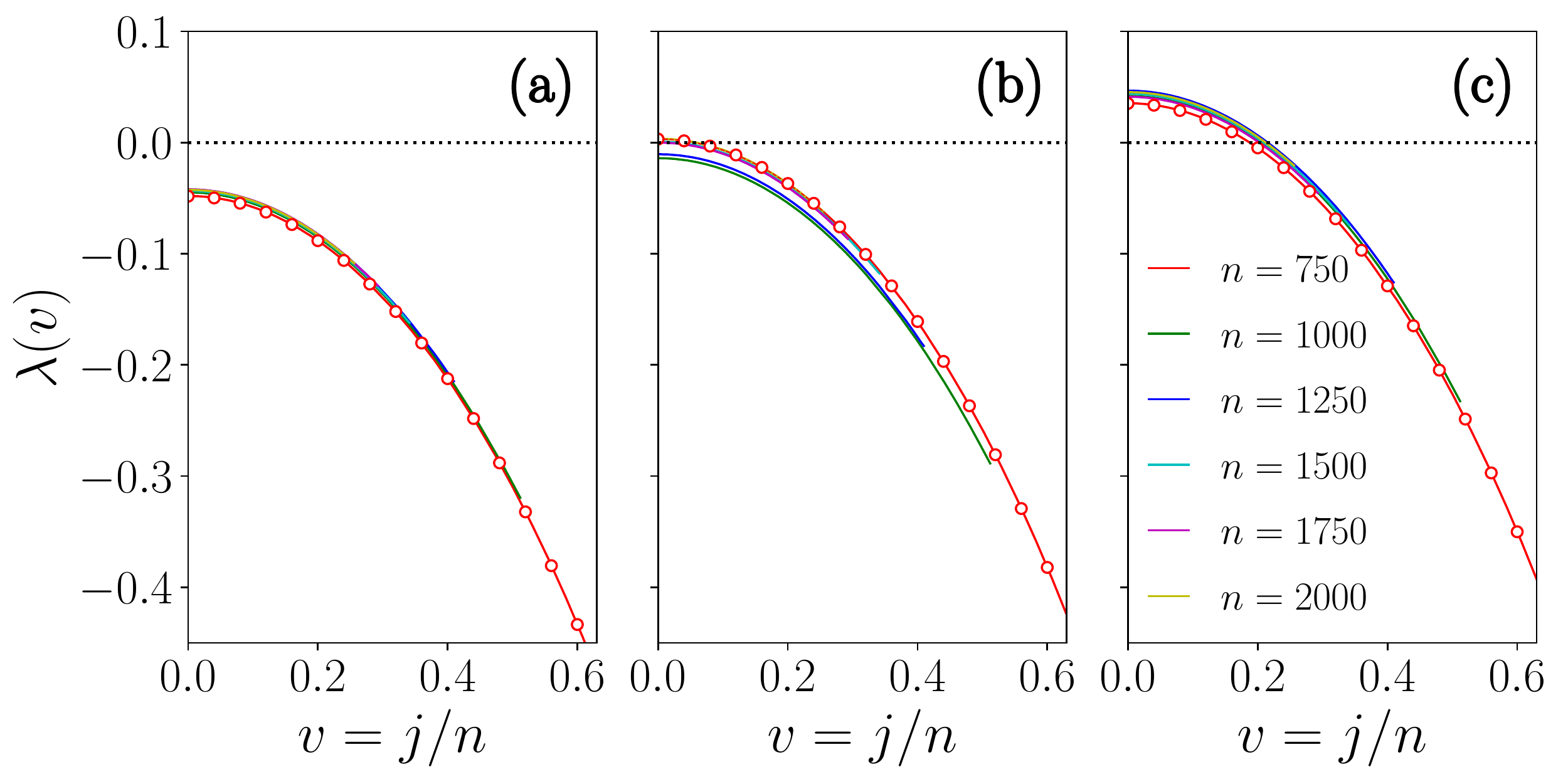}
\end{center}
\caption{Plots showing the dependence of FTLEs, $\lambda(v)$ in Eq.~\ref{eq:cml:le}, on the velocity $v = j/n$:  (a) $\alpha = 2.92$ (non-chaotic), (b) $\alpha = 3.04$ (strange non-chaotic) and (c) $\alpha = 3.10$ (chaotic). The other parameters are $\epsilon' = 0.76$ and $\kappa = 0.5$ with $\epsilon = \epsilon' (4/\alpha - 1)$. The circles correspond to the fit with Eq.~\ref{eq:fit} for $n=750$.}
\label{logis-lamv}
\end{figure}

Another way to understand the exponential spread or decay of the OTOC is to correlate the FTLEs with the velocity by calculating the velocity dependent Lyapunov exponents (VDLE), $\lambda(v)$, with $v = j/n$~\cite{Chatterjee2020}. The VDLE helps to visualise how the perturbation grows or decays in a frame moving with the velocity  $v = j/n$. To begin with, in Fig.~\ref{logis-lamv}(a) we plot $\lambda(v)$ as a function of $v$ for the case with $\alpha = 2.92$, $\epsilon' = 0.76$ and $\kappa = 0.5$. %
Next, we repeat the experiment for the SNA case with $\alpha = 3.04$, $\epsilon' = 0.76$ and $\kappa = 0.5$.
For the chaotic case, the plot $\lambda(v)$ against $v$ is shown in Fig.~\ref{logis-lamv}(c). In all the three cases, $\lambda(v)$ falls off with the increase in $v$ from an initial $\lambda'$. On examining the plots carefully, we identify that the variation of the VDLE with respect to the velocity can be represented by the form
\begin{align}
\lambda(v) = & \lambda' + \frac{1}{a} \left( 1- \cosh \frac{v}{b}  \right), \label{eq:fit}
\end{align}
where $\lambda'$, $a$ and $b$ are constants.  In Table \ref{tab:01} we present the fitting parameters, $\lambda'$, $a$ and $b$  for the three cases considered above for different values of $n$.
\begin{table}[!ht]
\caption{Estimate of the fitting parameters $\lambda'$, $a$ and $b$ according to Eq.~\ref{eq:fit} for the variation of VDLE with respect to $v = j/n$ in the CML of forced logistic maps for different $\alpha$ and $n$. The error in the estimates of $a$ and $b$ are less than or equal to $10^{-3}$, while the error in $\lambda'$ is of the order of $10^{-5}$ or less.}
\begin{center}
\begin{tabular}{crrrr|rrrr}
$\alpha$ &  \multicolumn{1}{c}{$n$} & \multicolumn{1}{c}{$\lambda'$} & \multicolumn{1}{c}{$a$} & \multicolumn{1}{c|}{$b$} & 
\multicolumn{1}{c}{$n$} & \multicolumn{1}{c}{$\lambda'$} & \multicolumn{1}{c}{$a$} & \multicolumn{1}{c}{$b$}  \\
\hline
$ 2.92$ & $ 750$ & $-0.048$ & $ 1.30$ & $ 0.623$ & $1500$ & $-0.044$ & $ 1.04$ & $ 0.693$ \\
$ 3.04$ &        & $ 0.003$ & $ 1.30$ & $ 0.623$ &        & $ 0.001$ & $ 1.04$ & $ 0.694$ \\
$ 3.10$ &        & $ 0.035$ & $ 1.30$ & $ 0.622$ &        & $ 0.044$ & $ 1.04$ & $ 0.692$ \\
\hline
$ 2.92$ & $1000$ & $-0.045$ & $ 1.14$ & $ 0.663$ & $1750$ & $-0.042$ & $ 1.01$ & $ 0.702$ \\
$ 3.04$ &        & $-0.014$ & $ 1.14$ & $ 0.663$ &        & $-0.000$ & $ 1.02$ & $ 0.700$\\
$ 3.10$ &        & $ 0.042$ & $ 1.14$ & $ 0.663$ &        & $ 0.041$ & $ 1.02$ & $ 0.701$\\
\hline
$ 2.92$ & $1250$ & $-0.042$ & $ 1.08$ & $ 0.681$ & $2000$ & $-0.042$ & $ 0.99$ & $ 0.710$ \\
$ 3.04$ &        & $-0.010$ & $ 1.08$ & $ 0.681$ &        & $ 0.003$ & $ 0.99$ & $ 0.711$\\
$ 3.10$ &        & $ 0.047$ & $ 1.08$ & $ 0.681$ &        & $ 0.045$ & $ 0.99$ & $ 0.710$\\
\hline
\end{tabular}
\end{center}
\label{tab:01}
\end{table}
Interestingly, for a given $n$, the values of $a$ and $b$ remain the same for all the three regions, namely non-chaotic, SNA and chaotic dynamics. Surprisingly, in the CML of quasiperiodically forced cubic maps (see Table~\ref{tab:02}), the variations of the VDLE with respect to the velocity can also be represented by the same expression \ref{eq:fit} and more intriguingly the fitting parameters $a$ and $b$ also coincide with the one found for the CML of forced logistic maps.  Further, we observe that certain continuous dissipative systems that exhibit SNA also follow a similar description as given in Eq.~\ref{eq:fit} for the VDLE. 

To summarize, we come across  the following observations in the study of VDLE  $\lambda(v)$  in spatially extended systems that admit SNA dynamics:
\begin{itemize}
\item[(i)] The dependence of $\lambda(v)$ on $v$ can be represented by a universal functional form which is valid for all the dynamical regimes. This property differs the one observed in Duffing chain where the functional form differs in each regime~\cite{Chatterjee2020}. 
\item[(ii)] The functional form given in Eq.~\ref{eq:fit} also holds good for a class of dynamical systems that exhibit SNA.
\end{itemize}
It is clear from above that the spatiotemporal spread in the CML of forced logistic maps in the SNA regime differs from that of the nonchaotic and chaotic regimes. To provide a broader perspective of the OTOC, we present a two parameter phase diagram in the $\alpha - \epsilon'$ plane by plotting the IS  $v_b(n, D_{th})$ and the Lyapunov exponent $\lambda_0(n)$  computed at $D_{th} = 1$ after $n=1200$ iterations with $M=1024$ ($2049$ lattice points). The choice of this size is to avoid the spread hitting the boundary in the chaotic regime and this does not alter the results. %
\begin{figure}[!ht]
\begin{center}
\includegraphics[width=\linewidth]{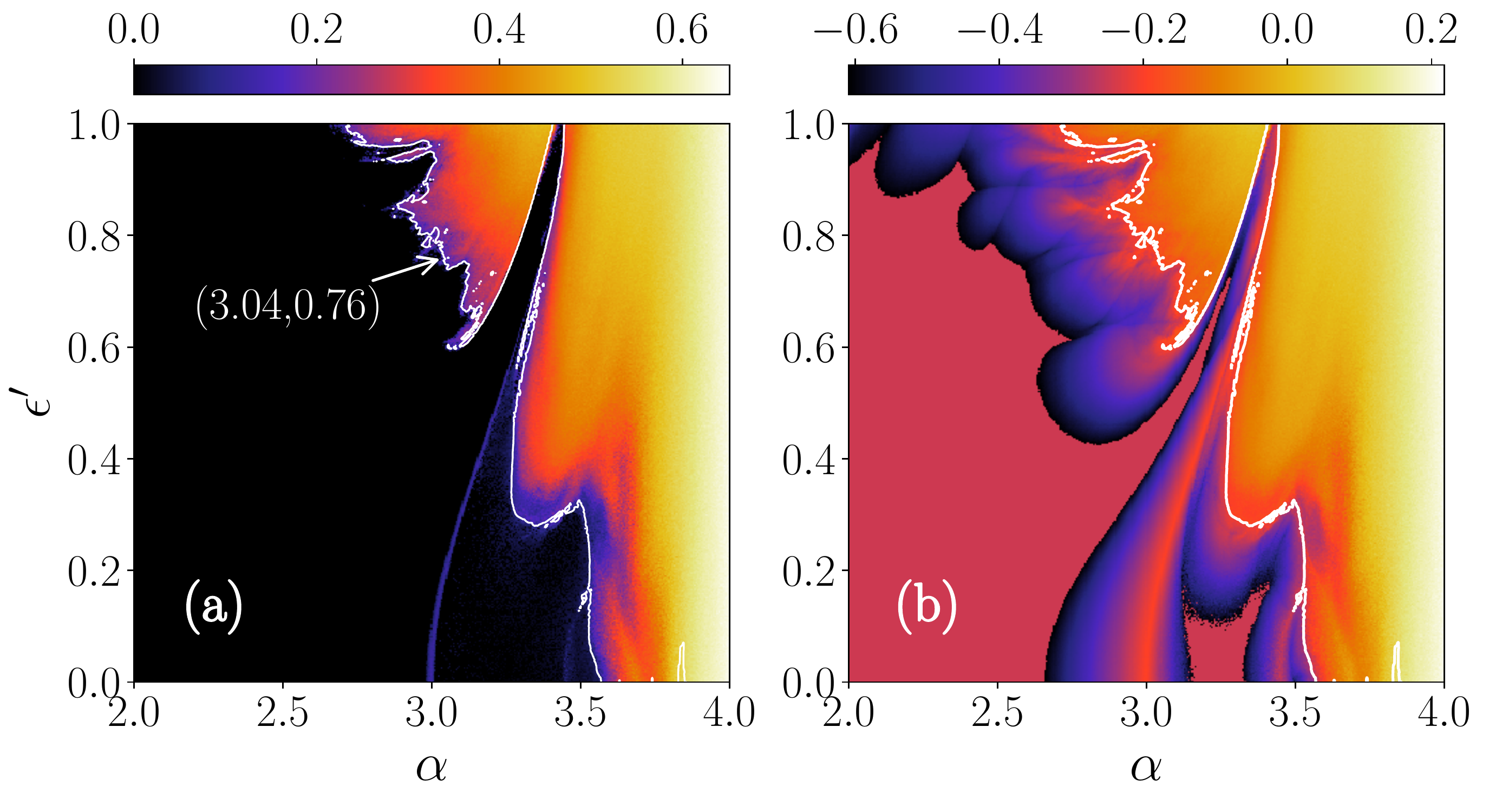}
\end{center}
\caption{Two parameter phase diagrams of the CML with quasiperiodically forced logistic maps depicting (a) the instantaneous speed $v_b$ and (b) FTLE $\lambda_0$ in the $\alpha - \epsilon'$ plane, where $\epsilon'  = \epsilon / (4/\alpha - 1) $ in a grid of size $401 \times 401$  after $1200$ iterations. The white line indicates $\lambda_{m} = 0$,  where $ \lambda_{m}$ is the largest long time Lyapunov exponent of the whole CML Eqs.~\ref{eq:cml:01}  and \ref{eq:cml:02} as calculated using Wolf et al. algorithm~\cite{Wolf1985}. The other parameters are $\Omega = (\sqrt{5}-1)/2$ and $\kappa = 0.5$.  The white arrow locates the point $(\alpha = 3.04, \epsilon' = 0.76)$ in the parameter space that is used to demonstrate the SNA nature in the CML of forced logistic maps.}
\label{logis-phase}
\end{figure}%
In Figs.~\ref{logis-phase}(a) and \ref{logis-phase}(b) we depict the phase diagrams of the IS $(v_b)$ and the Lyapunov exponent $(\lambda_0)$, respectively. In Fig.~\ref{logis-phase}, the white and black shades indicate the parameters at which $\lambda_0 = 0$. The regions with $v_b \approx 0.2$ exhibit the presence of SNA in the time series. The white arrow locating at $\alpha = 3.04, \epsilon' = 0.76$ in Fig.~\ref{logis-phase}(a), confirms the SNA nature in the CML of forced logistic maps.  Further, it may be noted that $v_b$ has a one-to-one correspondence with $\lambda_0$ and has its maximum value in the chaotic regime. These two phase diagrams have a close resemblance with that of the isolated map, see for example Refs.~\cite{Prasad1997, Prasad1998, Prasad2001}. In other words, the CML of forced logistic maps~\ref{eq:cml:01} behave qualitatively similar to that of the isolated logistic map \ref{eq:cml:02} and that the SNA nature is preserved in the CML as well. 
 
\subsection{Characterisation of SNA in the CML of quasiperiodically forced logistic maps}

The presence of SNA in the CML can be confirmed through a few quantitative measures such as, (i) the distribution of finite time Lyapunov exponents, (ii) the phase sensitivity, (iii) the partial Fourier sums and (iv) $0-1$ Test~\cite{Prasad2001,Venkatesan2001,Gopal2013,Toker2020}. To confirm that the heat map displayed in Fig.~\ref{sna-logis}(b) refers the SNA dynamics, in the following, we evaluate the above four measures for the CML of forced logistic maps~\ref{eq:cml:01}.

In the case of SNA, the largest Lyapunov exponent is negative, however, the distribution of FTLE helps one to examine the strangeness present in the time series. The stationary density of the FTLE  $P(N,\lambda) \, d\lambda$ is defined as the probability that the local Lyapunov exponent $\lambda_N$ takes a value between $\lambda$ and $\lambda + d\lambda$. $P(N,\lambda)$ can be estimated by dividing a long time series into segments of length $N$, and calculating the local Lyapunov exponent $\lambda_N$ in each segment. We look at the distribution of the largest FTLE in the CML of forced logistic maps~\ref{eq:cml:01} for the choice of parameters used in Fig.~\ref{sna-logis}. The largest FTLE can be computed as an average over a finite number of iterations, for example $n=200$, by following the standard procedure~\cite{Wolf1985}. In particular, we look at the distribution of $P(200, \lambda)$ for the choices $\alpha = 2.92$, $3.04$ and $3.10$ with the other parameters being fixed at $\epsilon' = 0.76$ and $\kappa = 0.5$.  For these choices, the isolated map exhibits, nonchaotic, SNA and chaotic type of dynamics, respectively. %
\begin{figure}[!ht]
\begin{center}
\includegraphics[width=\linewidth]{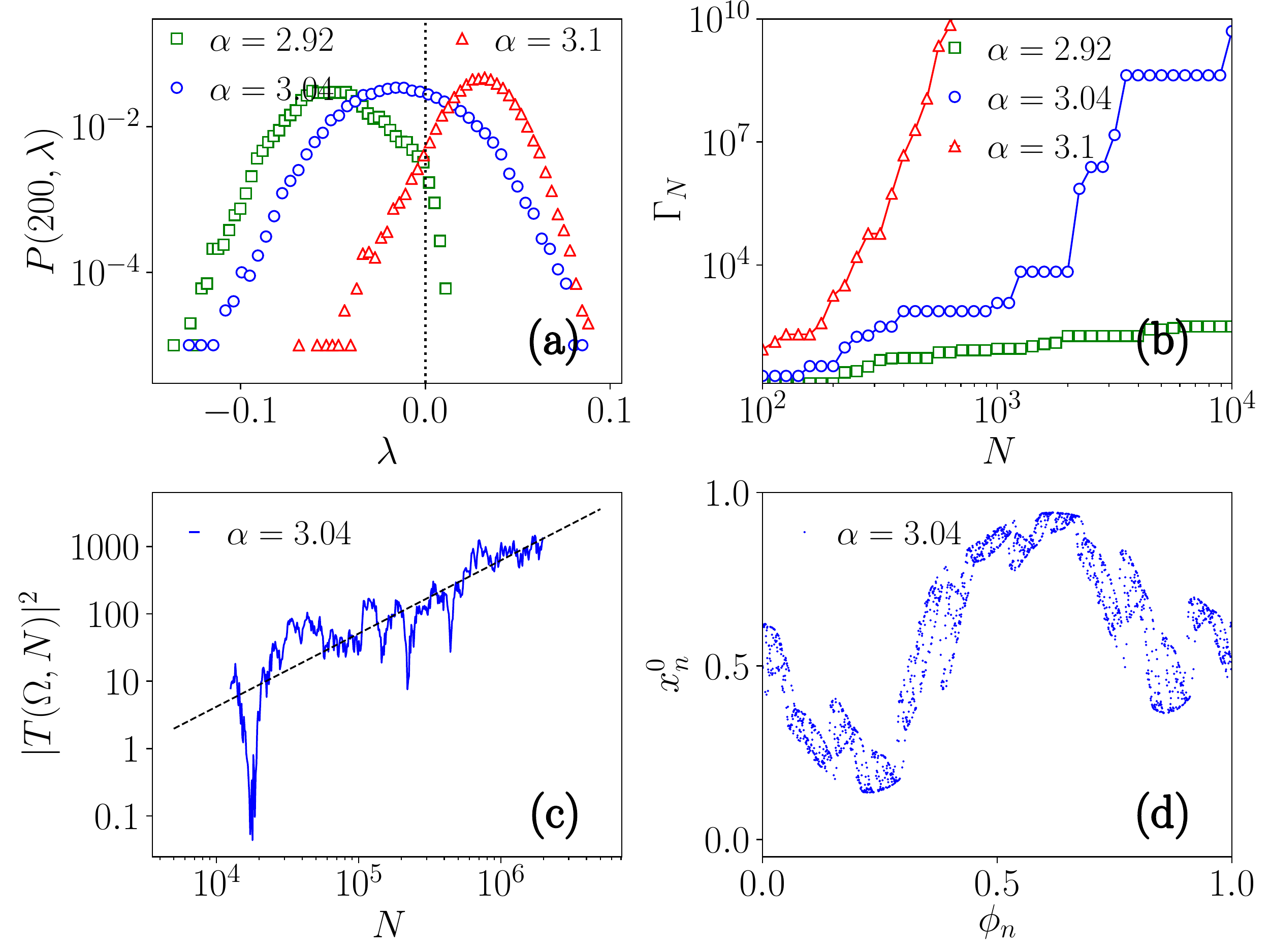}
\end{center}
\caption{Plots of the characteristic measures of SNA for the CML of forced logistic maps: (a) the distribution of finite time Lyapunov exponents (FTLE) $P(200,\lambda)$, (b) phase sensitivity function $\Gamma_N$ versus $N$, and (c) $\vert T(\Omega, N)\vert^2$ showing singular continuous spectrum, and (d) projection of the strange nonchaotic attractor in the $(x^{0} - \phi)$ plane showing SNA at $\alpha = 3.04$ and $\epsilon' = 0.76$.}
\label{logis-all}
\end{figure}%
In Fig.~\ref{logis-all}(a), we plot the distribution of the largest FTLE in the logarithmic scale for the above three cases. With $\alpha= 2.92$, the largest FTLE is distributed with a mean around a small negative as shown by green squares in Fig.~\ref{logis-all}(a). It further approaches towards zero for $\alpha = 3.04$, and is distributed about a mean negative as illustrated with blue circles in the same figure. This kind of distribution is a typical signature for the presence of SNA~\cite{Pikovsky1995a,Prasad1997}. In the chaotic regime, the FTLE is distributed around a finite positive value as shown by the red triangles in Fig.~\ref{logis-all}(a).

The presence of SNA can also be confirmed through the phase sensitivity exponent~\cite{Prasad2001,Venkatesan2001}. It can be evaluated by detecting the non-differentiability of the time series $x(\phi)$ when viewed as a fractal curve and by examining the separation of points that are initially close to $\phi$~\cite{Pikovsky1995,Prasad2001,Venkatesan2001}. This measure of strangeness can be determined by calculating the derivative $dx/d\phi$ along an orbit and finding its maximum value as the smallest of such realization for arbitrary $(x,\phi)$.  The phase density function $\Gamma_N$ is defined by~\cite{Pikovsky1995,Prasad2001}
\begin{align}
\Gamma_N = \text{min}_{x,\phi} \left[ \text{max}_{1<N} \left\vert \frac{dx_N}{d\phi} \right\vert \right].
\end{align}
For a chaotic time series, the sensitivity grows exponentially while it shows power law dependence, i.e. $\Gamma_N \sim N^\beta$, for SNA. However,  $\Gamma_N$ diverges for very large $N$~\cite{Prasad2001}. 

We examine the nature of $\Gamma_N$ in the CML of forced logistic maps \ref{eq:cml:01}. In Fig.~\ref{logis-all}(b) we depict the variation of $\Gamma_N$ with respect to $N$ for the same three parametric choices considered above, namely $\alpha = 2.92$, $3.04$ and $3.10$,  with $\epsilon' = 0.76$, $\omega = (\sqrt{5}-1)/2$, and $\kappa = 0.5$.  The green-squares, blue-circles and red-triangles in Fig.~\ref{logis-all}(b) represent $\Gamma_N$ for $\alpha = 2.92$, $3.04$ and $3.10$, respectively. We compute $\Gamma_N$ by considering the time series $x_n^j$ with $j=0$, however, the results are robust for $j \neq 0$. The SNA nature of the CML is evident as $\Gamma_N \sim N^\mu$ (power-law relation) for $\alpha = 3.04$. The exponential growth of $\Gamma_N$  with respect to $N$ as indicated by red-triangles in Fig.~\ref{logis-all}(b confirms the chaotic case.

Yet another quantitative distinction of the SNA from the other two types of dynamics may be provided through the discrete Fourier transform of the time series $\{ x_n^j\}$. For instance, the partial Fourier sums, defined as~\cite{Pikovsky1995,Yalcinkaya1997,Prasad2001}
\begin{align}
T(\Omega, N) = \sum_{k=1}^{N} x_k \exp\left( \mathrm{i} 2\pi k \Omega  \right), \label{eq:fsum}
\end{align}
where $\Omega$ is proportional to the irrational driving frequency $\omega$, can be used to describe the SNA nature. It has been shown that SNA exhibits singular-continuous spectrum, i.e. $\vert  T(\Omega, N)  \vert^2 \sim N^\beta$ with $1 < \beta < 2$. In Fig.~\ref{logis-all}(c), we show the spectrum of $\vert  T(\Omega, N)  \vert^2$ with respect to $N$  from the time series $\{ x_n^0\}$ for the CML \ref{eq:cml:01} with $\alpha = 3.04$. The outcome confirms the singular continuous spectrum exhibited by the time series with  $\beta \approx 1.09$. From Fig.~\ref{logis-all}(d), we can visualise the projection of the strange nonchaotic attractor in the $x^0 - \phi$ plane for $\alpha = 3.04$. This attractor is similar to the one in which SNA created through fractalization described for isolated logistic map~\cite{Prasad2001}.

The $0-1$ test helps to  distinguish periodic (nonchaotic - limit cycles, quasiperiodic orbits, etc.), SNA and chaotic attractor, from the time series~\cite{Gopal2013,Toker2020}. This will produce a number $0$ for torus, $1$ for chaos, and intermediate value between $0$ and $1$ for SNA~\cite{Gopal2013}. For a given time series $x_i, \, i = 1,2, \ldots, N$, we define a set of two translational variables $p(n)$ and $q(n)$ as~\cite{Gottwald2009} 
\begin{align}
p(n) = \sum_{k=1}^{n} x_k\cos c k, \;\;\; q(n) = \sum_{k=1}^{n} x_k \sin c k, \label{eq:z1:01}
\end{align}
where $c \in (0,\pi)$. The diffusive (or non-diffusive) behaviour of $p$ and $q$ can be investigated by analyzing the mean squared displacement $\mathcal M$, which is defined by  
\begin{align}
{\mathcal M}(l) = \lim_{N \to \infty}  \frac{1}{N-l} \sum_{k=1}^{N-l}  & \left( \left[p(k+l) - p(k) \right]^2  \right . \notag \\ &
\left. +   \left[q(k+l) - q(k) \right]^2  \right), \label{eq:z1:02}
\end{align}
where $l = 1,2,\ldots,  l_{max}$ with $l_{max}$ usually taken as $N/10$. 
If the dynamics is regular then the mean square displacement is a bounded function in time, whereas for chaos the mean squared displacement scales linearly with time~\cite{Gopal2013}. 

A linear regression for the log-log plot of the mean squared displacement, defined as
\begin{align}
{\mathcal K} = \lim_{l \to \infty} \frac{\log {\mathcal M}(l)}{\log l} \label{eq:z1:03},
\end{align}
quantifies the $0-1$ test. %
Alternatively, $\mathcal K$ can be computed from the correlation through 
\begin{align}
{\mathcal K}= \mbox{corr}(\xi, {\mathcal M}) = \frac{\mbox{cov}(\xi, {\mathcal M})}{\sqrt{\mbox{var}(\xi) \mbox{var}({\mathcal M})}}, \label{eq:k:covar}
\end{align}
where $\mbox{cov}(\cdots)$ corresponds to covariance, $\mbox{var}(\cdots)$ represents the variance, $\xi = 1,2,\ldots, l_{max}$ and ${\mathcal M} = \{ {\mathcal M}(1), {\mathcal M}(2), \ldots,  {\mathcal M}(l_{max}) \}$.

To employ the $0-1$ test, we generate a time series $\{x_i^j,\, \, i = 1, 2, \ldots, N\}$ with $j=0$ of length $N=10000$ by iterating Eqs.~\ref{eq:cml:01} and \ref{eq:cml:02}, and compute $\mathcal K$ from Eq.~\ref{eq:k:covar} as described above. For the choice $\alpha = 2.92$, $3.04$ and $3.10$,  $\mathcal K$ takes the values $0.14$, $0.68$, and $0.96$, respectively. Thus the $0-1$ test also confirms the SNA nature of the times series for $\alpha = 3.04$. 

The $0-1$ test can be extended to identify the regions of the parameters for which the CML of forced logistic maps Eqs.~\ref{eq:cml:01} and \ref{eq:cml:02} exhibit nonchaotic, SNA and chaotic dynamics. %
\begin{figure}[!ht]
\begin{center}
\includegraphics[width=0.99\linewidth]{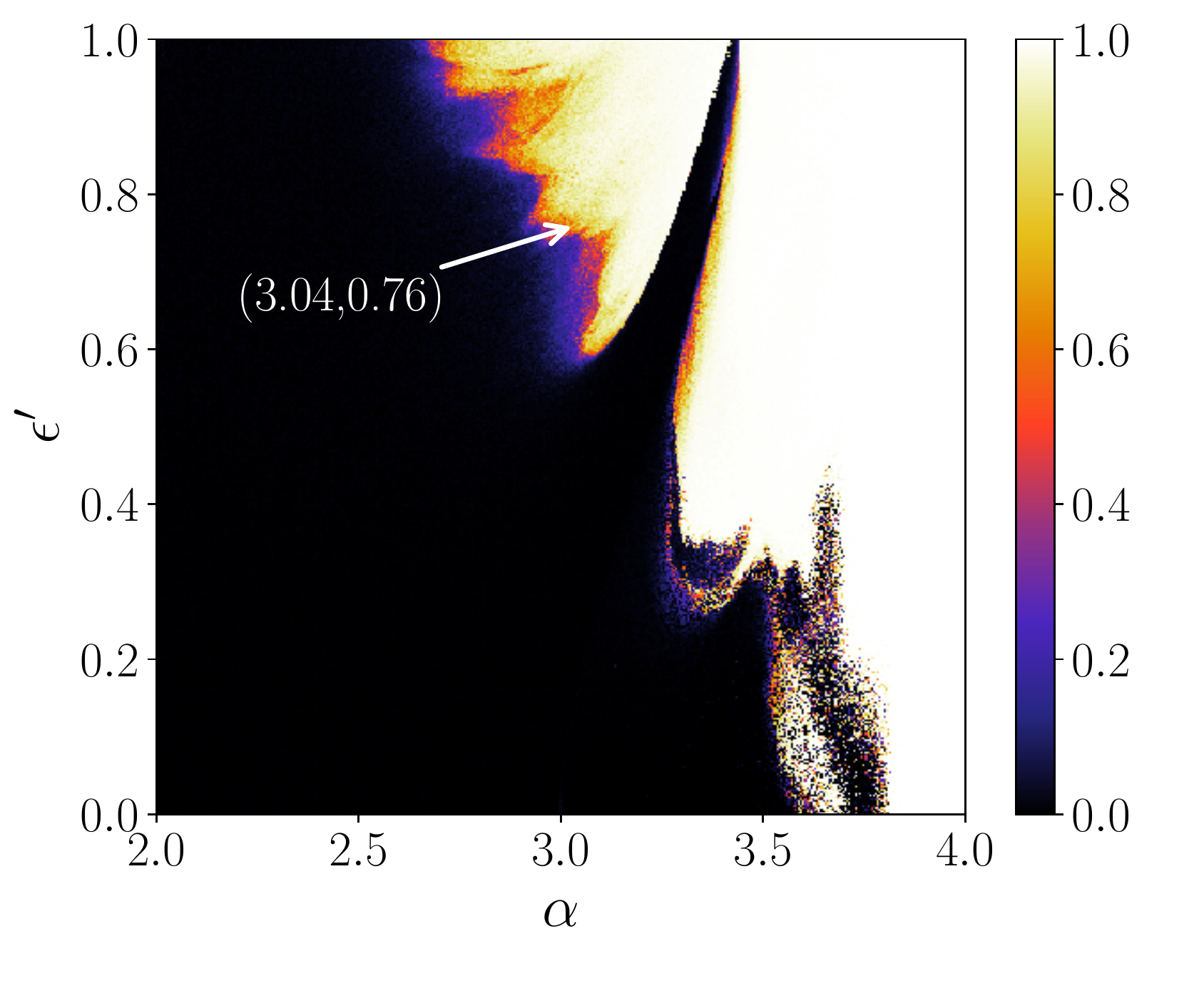}
\end{center}
\caption{Two parameter phase diagrams of the CML with quasiperiodically forced logistic maps \ref{eq:cml:01} and  \ref{eq:cml:02} showing density contours (heat map) of $\mathcal K$ \ref{eq:z1:03}. SNA is identified in the regions of $\mathcal K \in (0.2,0.8)$. The white arrow points the location of $\alpha = 3.04$, $\epsilon' = 0.76$.}
\label{fig-z1test}
\end{figure}%
We extend the calculation of $\mathcal K$ for every pair of $(\alpha, \epsilon')$, where $\alpha \in (2,4)$ and $\epsilon' \in (0,1)$, in a grid of size $421 \times 421$.
Fig.~\ref{fig-z1test} depicts a two-parameter phase diagram of  $\mathcal K$ in the $\alpha - \epsilon'$ plane. In Fig.~\ref{fig-z1test}, the regions where $\mathcal K \in (0.2,0.8)$ identify the parameters $(\alpha, \epsilon')$ for which SNA can be found. The white arrow spotted at the location $\alpha = 3.04$, $\epsilon' = 0.76$ confirms SNA in the CML of forced logistic maps.

Above,  we have examined  the IS and  the FTLE in different dynamical regimes of the CML. The spread of the OTOC in the nonchaotic and chaotic regime of the forced logistic map behaves as expected. That is, in the nonchaotic case, the initial spread ceases after a finite number of iterations, and it shows ballistic spread with light-cone like structure when the CML exhibits spatiotemporal chaos. However, a distinct on and off kind of spread is identified in the case of SNA. Further, the presence of SNA in the CML is established through the analysis of the distribution of FTLEs, phase sensitivity, partial Fourier sums, and $0-1$ test. 

\section{Coupled map lattices of quasiperiodically forced  cubic maps}
\label{sec:cml:2}

In the previous section, we studied the nature of OTOC in the CML of quasiperiodically forced logistic maps and  observed that the on and off kind of spread that is present in the SNA regime differs from the nonchaotic and chaotic regimes. It is vital  to confirm whether, the other dynamical systems that exhibit SNA also produce the same OTOC pattern or not. For this purpose, we consider another well-studied map in the literature, namely the quasiperiodically driven cubic map which has a similar form of \ref{eq:cml:01} and is described by~\cite{Venkatesan2001,Gopal2013}.
\begin{subequations}
\label{eq:ccml:01}
\begin{align}
& x_{n+1} = Q + F   \cos (2 \pi \phi_n ) - A  x_n +  x_n ^3 = f(x_n), \label{eq:ccml:01a} \\
& \phi_{n+1} = \phi_n + \omega \; (\text{mod\ } 1), \label{eq:ccml:01b}
\end{align}
\end{subequations}%
where $\omega = \left( \sqrt{5} - 1\right) /2$ is the irrational driving frequency,  $F$ represents the forcing amplitude, and $Q$ and $A$ are the other control parameters.  The  cubic map \ref{eq:ccml:01} has a close analogy to the typical Duffing oscillator and exhibits SNA for a wider range of parameters~\cite{Venkatesan2001, Gopal2013}. 

\subsection{CML of quasiperiodically forced cubic maps: OTOC}

We analyse the spatial spread of the OTOC in the CML of cubic maps~\ref{eq:ccml:01} from the linearized CML of the form \ref{eq:cml:lin01} with $f'(x)$ given by
\begin{align}
f'(x) =  3  x^2  - A.
\label{eq:cml:lin02}
\end{align}
We fix the parameters as $Q=0$, $F = 0.6$, $\kappa = 0.5$ and study the OTOC by varying the parameter $A$. For this choice of parameters, the isolated map \ref{eq:ccml:01} traverse from quasiperiodic attractor to chaotic through SNA when $A$ is varied in the range $(0.8, 2.4)$~\cite{Venkatesan2001, Gopal2013}.  The choice of parameters is arbitrary, however, we shall later show a complete phase diagram in the $F-A$ parameter plane, which confirms the aforesaid dynamics. We note that instead of the linearized map \ref{eq:cml:lin01} along with \ref{eq:cml:lin02} one may also consider two identical CMLs with an infinitesimal difference in the initial conditions at one lattice point. %
\begin{figure}[!htb]
\begin{center}
\includegraphics[width=\linewidth]{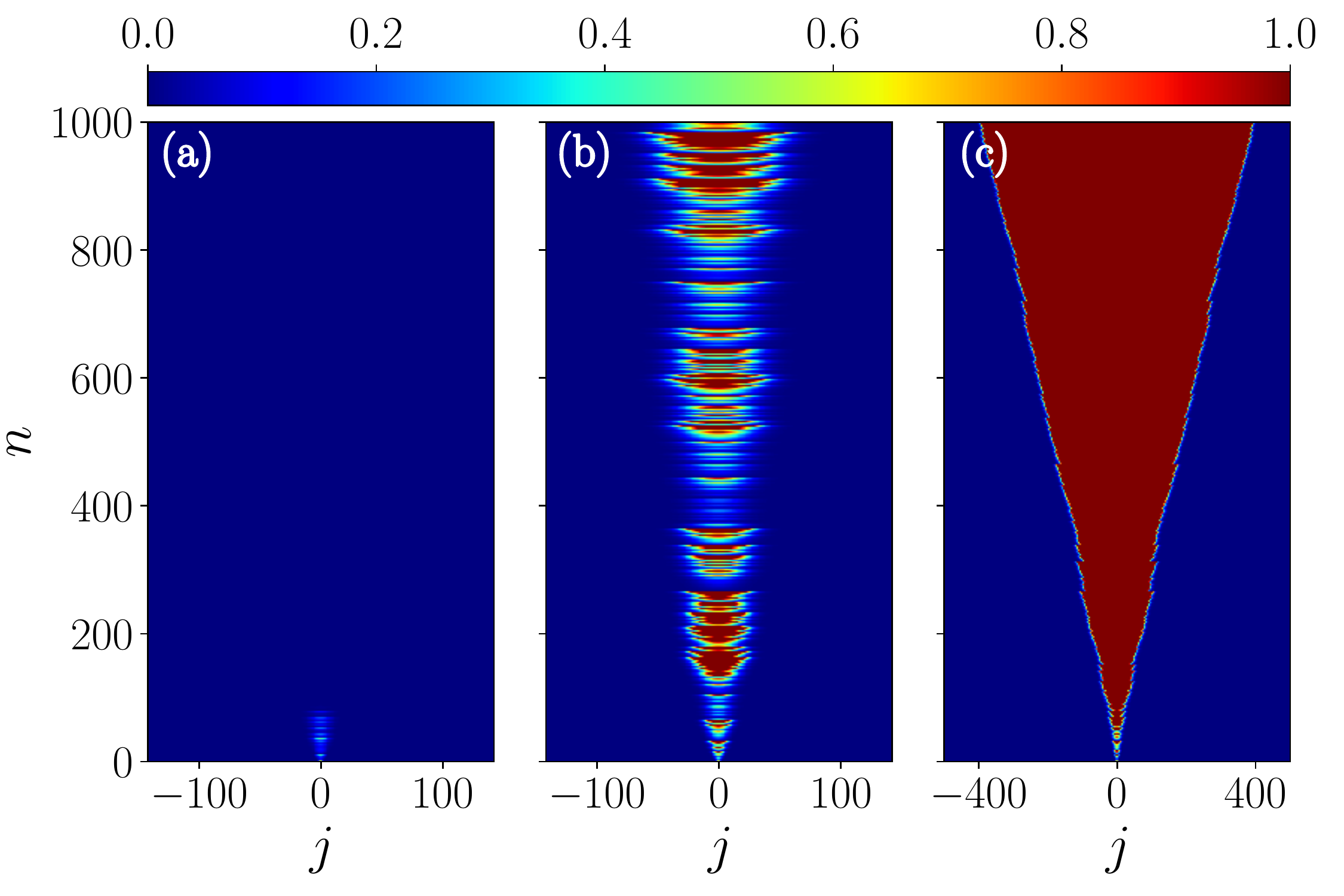}
\end{center}
\caption{Plot of the spatiotemporal spread of the perturbation in a CML of forced cubic maps in different dynamic regimes:  CML (Cubic map): (a) $A = 1.74$ (nonchaotic/periodic attractor), (b) $A = 1.7543$ (SNA) and (c) $A = 1.77$ (chaotic attractor). The other parameters are $Q = 0$, $F = 0.6$ and $\kappa = 0.5$.}
\label{sna-cubic}
\end{figure}%

\begin{figure}[!htb]%
\begin{center}
\includegraphics[width=\linewidth]{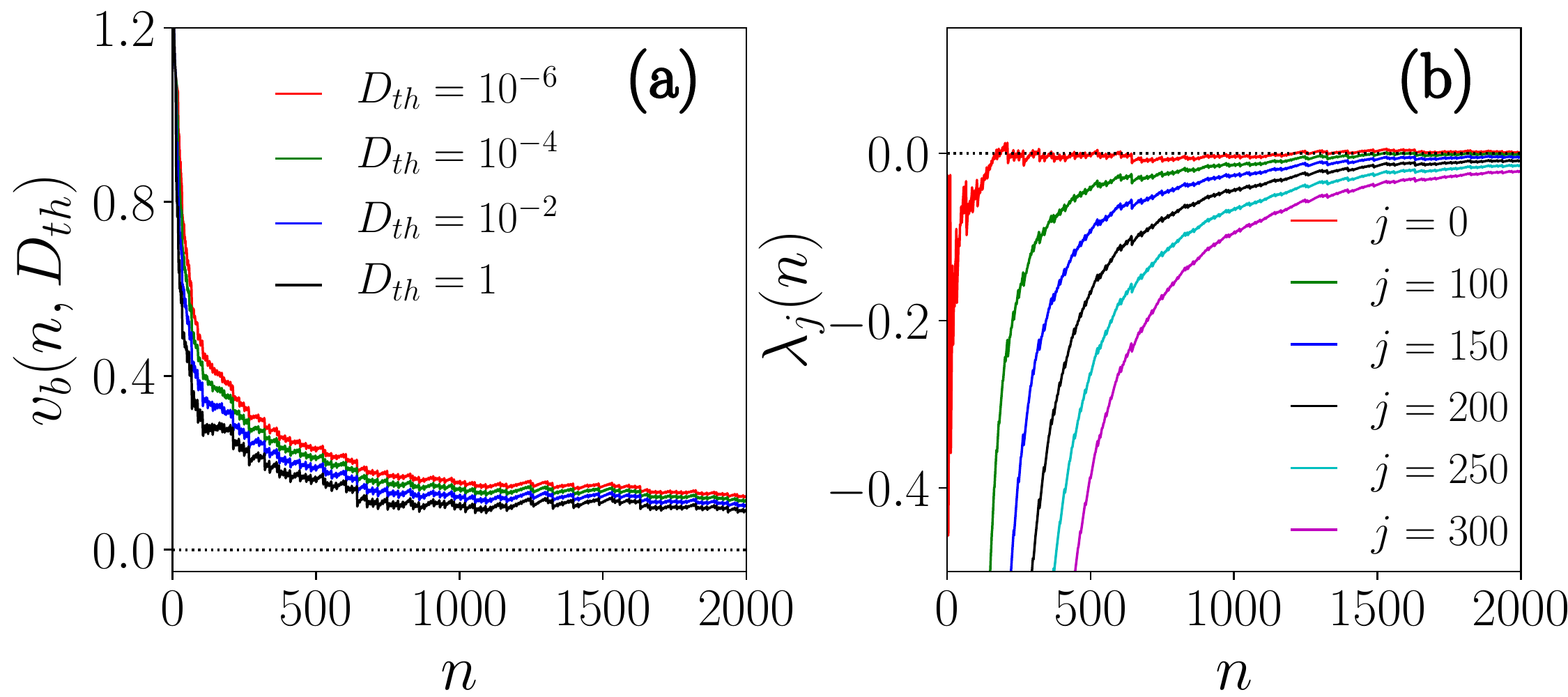}
\end{center}
\caption{CML (Cubic map): (a) Finite time Lyapunov exponent (FTLE) for different lattice points and (b) Instantaneous speed (IS) in the case of SNA. The parameters are $Q = 0$, $A =1.7543$, $F = 0.6$ and $\kappa = 0.5$.}
\label{cubic-lv-2}
\end{figure}

\begin{figure}[!htb] %
\begin{center}
\includegraphics[width=\linewidth]{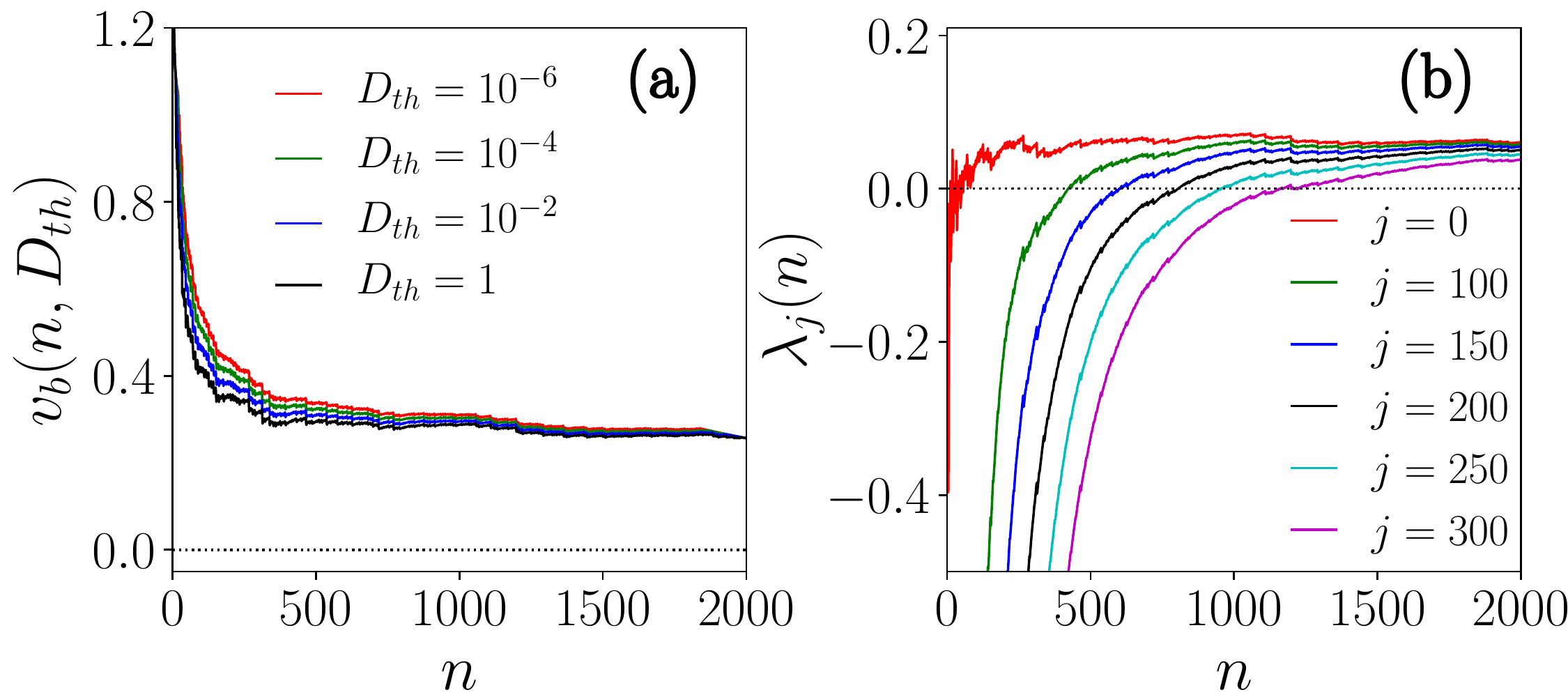}
\end{center}
\caption{CML (Cubic map): (a) Instantaneous speed (IS)  and   (b) finite time Lyapunov exponent (FTLE) for different lattice points in the case of chaotic attractor. The parameters are $Q = 0$, $A = 1.77$, $F = 0.6$ and $\kappa = 0.5$.}
\label{cubic-lv-3}
\end{figure}

In Fig.~\ref{sna-cubic}, we present the density contours (heat map) of the spatiotemporal spread of the OTOC \ref{eq:cml:otoc} for three different values of $A$. Fig.~\ref{sna-cubic}(a) illustrates the spread of the OTOC for $A = 1.74$ for which the isolated map \ref{eq:ccml:01} reveals nonchaotic (regular) dynamics. In this case, the initial spread of the OTOC is completely suppressed within fewer number of iterations which in turn confirms the nonchaotic nature of the cubic map \ref{eq:ccml:01} at the spatiotemporal scale.  Figure~\ref{sna-cubic}(b) displays the heat map of $D(j,n)$ for $A = 1.7543$ where the isolated map \ref{eq:ccml:01} exhibits SNA. The spread of the OTOC is similar to the one observed in the case of forced logistic maps discussed earlier. We observe intriguingly the same kind of repeated spread and suppression of the OTOC in the SNA regime. Here also the OTOC initially spreads to some lattices, then gets suppressed, again spreads and this occurrence continues. %
\begin{figure}[!htb]
\begin{center}
\includegraphics[width=\linewidth]{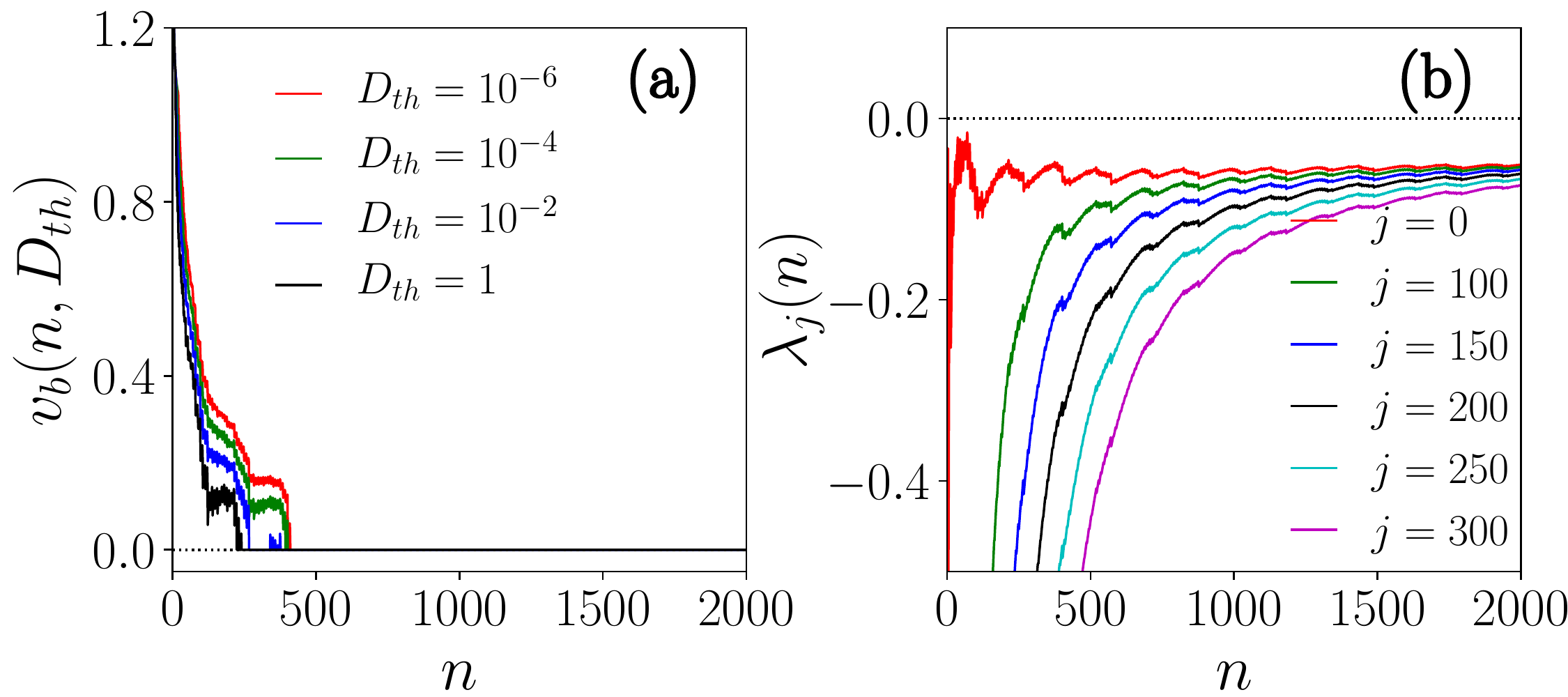}
\end{center}
\caption{CML (Cubic map): (a) Finite time Lyapunov exponent (FTLE) for different lattice points and (b) Instantaneous speed (IS) in the case of non-chaotic attractor. The parameters are $Q = 0$, $A = 1.74$, $F = 0.6$ and $\kappa = 0.5$.}
\label{cubic-lv-1}
\end{figure} %
On the other hand, in the chaotic regime ($A = 1.77$), as expected, the OTOC shows a ballistic spread as shown in Fig.~\ref{sna-cubic}(c). The heat map of $D(j,n)$ exhibits a light-cone like structure with sharp boundaries unlike the other two cases. We also compute the IS $v_b(n, D_{th})$, and FTLE $\lambda_j(x)$, of all three cases and depict the same in Figs.~\ref{cubic-lv-1} - \ref{cubic-lv-3}. In Fig.~\ref{cubic-lv-1}(a), we show the IS as a function of $n$ for different thresholds $D_{th}$, for $A = 2.74$, which drops to zero after fewer number of iterations. In Fig.~\ref{cubic-lv-1}(b), we plot the FTLE $\lambda_j(n)$ for $j \in (0, 100, 150, 200, 250, 300)$. %
As $n$ increases the largest FTLE converges to a steady value $(\approx -0.05)$ below zero, which indicates the nonchaotic nature.  For $A = 1.7543$, $v_b(n, D_{th})$ and $\lambda_j(n)$ evolve as shown in Fig.~\ref{cubic-lv-2}.  In this case, the speed $v_b$ stays near zero 
and the FTLEs remain negative in the vicinity of zero with the largest being of the order of $-0.001$. %
Finally, in the chaotic regime $(A = 1.77)$,  $v_b$ approaches to a steady value of $0.25$ and the  FTLEs approach a finite positive value as illustrated in Fig.~\ref{cubic-lv-3}.

\begin{figure}[!ht]
\begin{center}
\includegraphics[width=\linewidth]{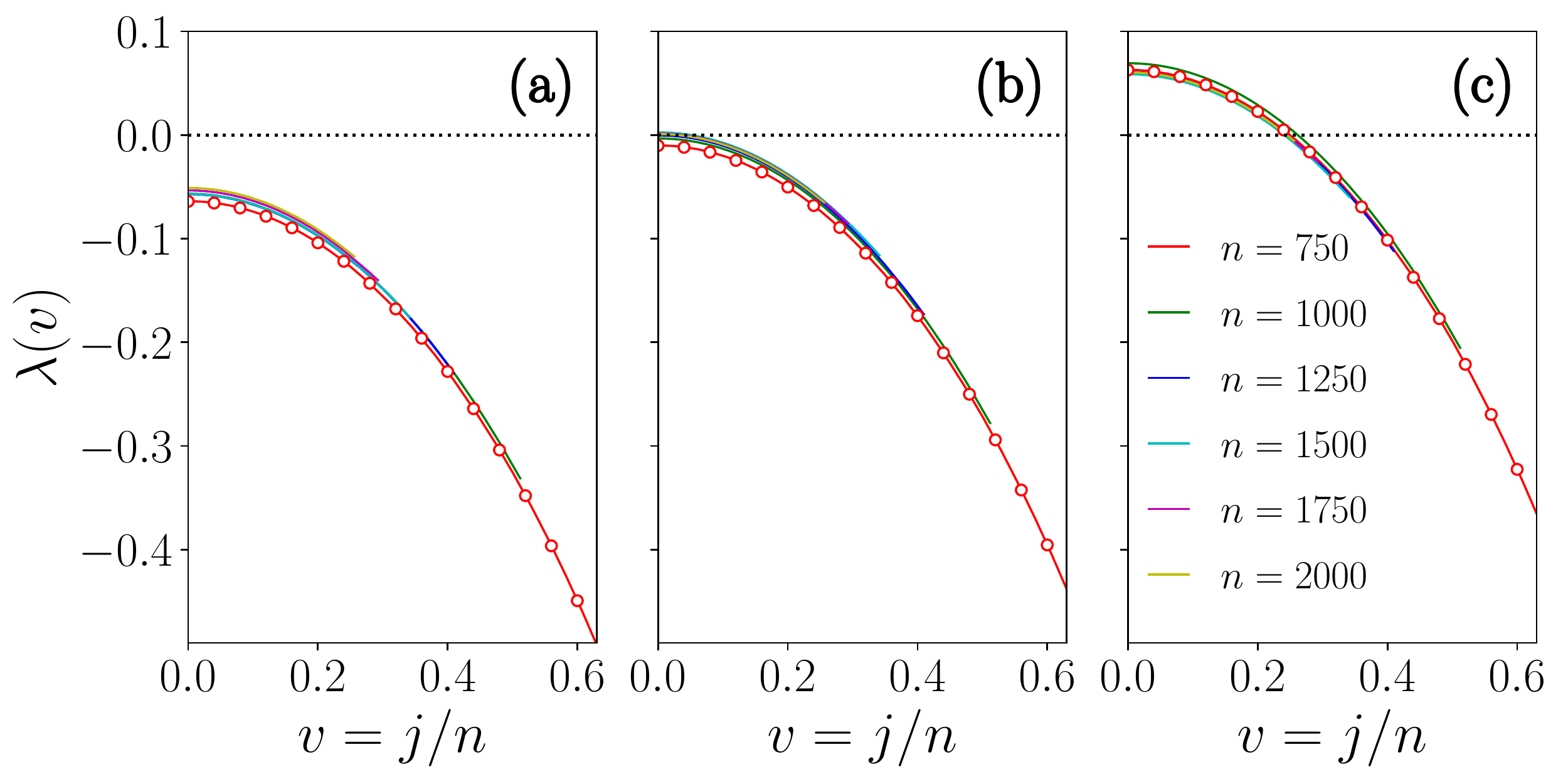}
\end{center}
\caption{Plots showing the dependence of FTLEs on the velocity: (a) $A = 1.74$ (nonchaotic/periodic attractor), (b) $A = 1.7543$ (SNA) and (c) $A = 1.77$ (chaotic attractor). The other parameters are $Q = 0$, $F = 0.6$ and $\kappa = 0.5$. The red-circles indicate the fit using Eq.~\ref{eq:fit} for $n=750$.}
\label{cubic-lamv}
\end{figure}

Next, we study the variation of VDLEs with respect to $j/n$ as before. In Fig.~\ref{cubic-lamv}, we show the plot of VDLEs versus the velocity in the  CML of quasiperiodically forced cubic maps for the values  $A = 1.74$, $1.7543$ and $1.77$.   %
The variations in VDLEs with respect to the velocity can be seen exactly in the same form as that of the CML of quasiperiodically forced logistic maps (see Fig.~\ref{logis-lamv} for a comparison). Surprisingly, the fit of $\lambda(v)$ also obeys Eq.~\ref{eq:fit}. Upon looking the spatiotemporal spread in the CML with the maps \ref{eq:cml:02} and \ref{eq:ccml:01} it is clear that the OTOC that comes out from the SNA regime is similar to each other. To our surprise the fitting parameters $a$ and $b$  found for the CML of cubic maps (Table~\ref{tab:02}) also match with the one obtained for the logistic maps (Table~\ref{tab:01}).  %
\begin{table}[!ht]
\caption{Estimate of the fitting parameters $\lambda'$, $a$ and $b$ according to Eq.~\ref{eq:fit} for the variation of VDLE with respect to $v = j/n$ in  the CML of forced cubic maps for different $A$ and $n$. The error in the estimates of $a$ and $b$ are less than or equal to $10^{-3}$, while the error in $\lambda'$ is of the order of $10^{-5}$ or less.}
\begin{center}
\begin{tabular}{lrrrr|rrrr}
 \multicolumn{1}{c}{$A$} &  \multicolumn{1}{c}{$n$} & \multicolumn{1}{c}{$\lambda'$} & \multicolumn{1}{c}{$a$} & \multicolumn{1}{c|}{$b$} & 
\multicolumn{1}{c}{$n$} & \multicolumn{1}{c}{$\lambda'$} & \multicolumn{1}{c}{$a$} & \multicolumn{1}{c}{$b$}  \\
\hline
$  1.74$ & $ 750$ & $-0.064$ & $ 1.30$ & $ 0.623$ & $1500$ & $-0.057$ & $ 1.04$ & $ 0.694$ \\
$1.7543$ &        & $-0.010$ & $ 1.30$ & $ 0.622$ &        & $ 0.003$ & $ 1.05$ & $ 0.691$ \\
$  1.77$ &        & $ 0.063$ & $ 1.30$ & $ 0.622$ &        & $ 0.059$ & $ 1.05$ & $ 0.692$ \\
\hline
$  1.74$ & $1000$ & $-0.057$ & $ 1.14$ & $ 0.663$ & $1750$ & $-0.053$ & $ 1.02$ & $ 0.701$ \\
$1.7543$ &        & $-0.003$ & $ 1.14$ & $ 0.663$ &        & $ 0.001$ & $ 1.01$ & $ 0.702$ \\
$  1.77$ &        & $ 0.069$ & $ 1.14$ & $ 0.663$ &        & $ 0.061$ & $ 1.02$ & $ 0.701$ \\
\hline
$  1.74$ & $1250$ & $-0.057$ & $ 1.08$ & $ 0.681$ & $2000$ & $-0.051$ & $ 1.00$ & $ 0.706$\\
$1.7543$ &        & $-0.007$ & $ 1.08$ & $ 0.682$ &        & $ 0.001$ & $ 0.99$ & $ 0.711$ \\
$  1.77$ &        & $ 0.061$ & $ 1.08$ & $ 0.682$ &        & $ 0.060$ & $ 0.99$ & $ 0.711$ \\
\hline
\end{tabular}
\end{center}
\label{tab:02}
\end{table}%
We believe that this coincidence is not arbitrary, and arise due to the structure of the CMLs and their dynamics as the isolated maps in both cases exhibit SNA. %
\begin{figure}[!ht]
\begin{center}
\includegraphics[width=\linewidth]{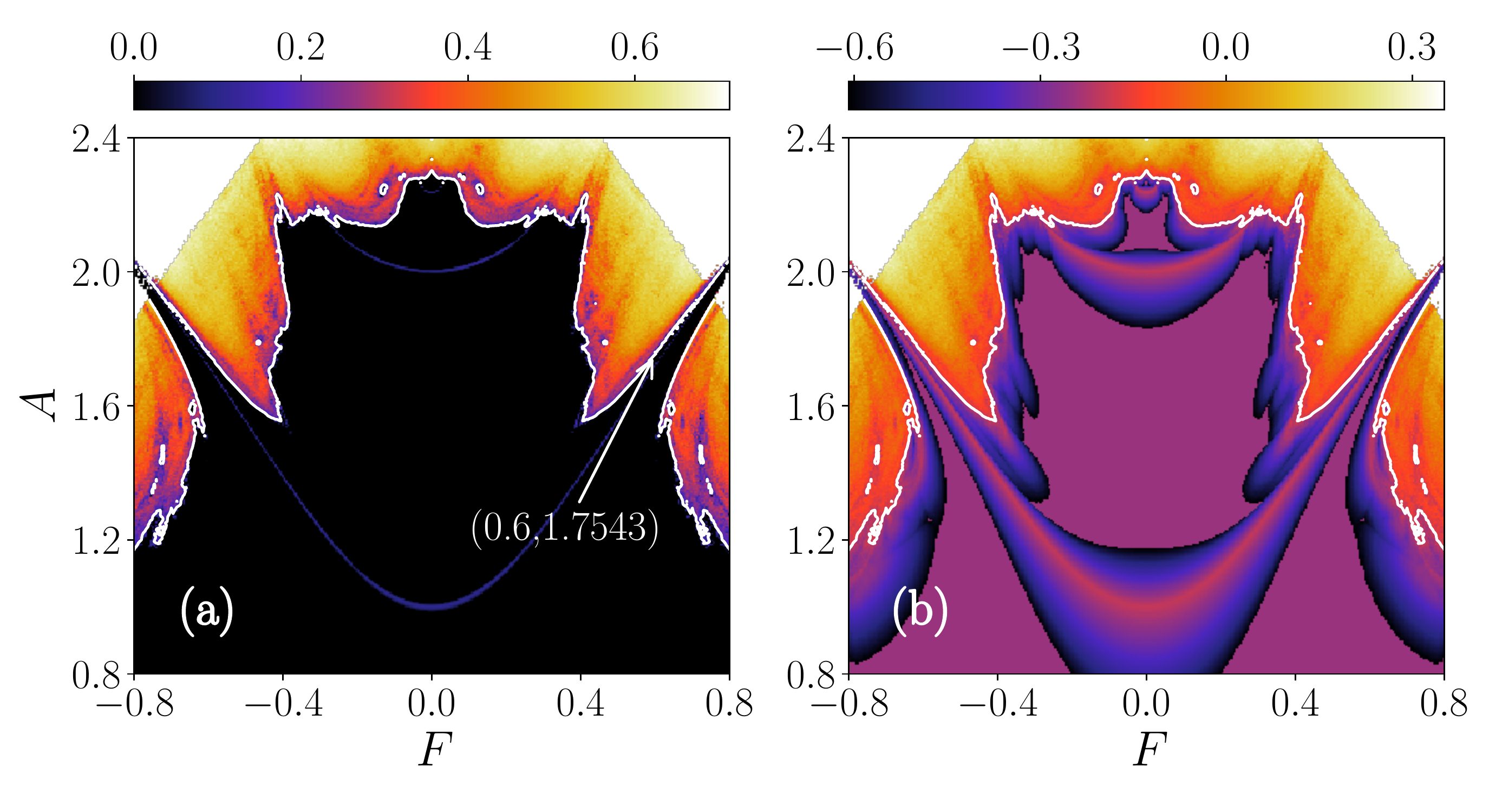}
\end{center}
\caption{Phase diagram for the CML of forced cubic maps showing (a) the instantaneous speed $v_b$ and (b) FTLE $\lambda_0$ as a function of $F$ and the parameter $A$ in a grid of size $321 \times 321$ after $1200$ iterations. The other parameters are $Q = 0$, $\Omega = (\sqrt{5}-1)/2$ and $\kappa = 0.5$. The white spaces near the top-left and top-right corners are the regions where the CML of cubic maps become unbounded. The white lines (contours) indicate $\lambda_m = 0$, with $ \lambda_m$ being the largest long time Lyapunov exponent of the whole CML Eqs.~\ref{eq:cml:01} and \ref{eq:ccml:01} as calculated using Wolf et al. algorithm~\cite{Wolf1985}. The white arrow identifies the point $(0.6, 1.7543)$ used to demonstrate the SNA in the CML.}
\label{cubic-phase}
\end{figure}%
We also examined the spread of OTOC in the continuous systems that exhibit SNA and found a similar behaviour of the OTOC and VDLEs in most of the cases. Though we have not investigated different routes to SNA in the CML, which requires a more detailed study, most of the dynamical features of the isolated systems are reflected in the spatiotemporal scale. We highlight that the peculiar on and off spread of OTOC that can be seen in the CMLs are robust and generic for systems showing SNA.  

We also present a two parameter phase diagram in the $F-A$ plane by plotting the IS,  $v_b(n,D_{th})$ and the Lyapunov exponent, $\lambda_0(n)$  computed at $D_{th} = 1$ after $n=1200$ iterations with $M=1024$ ($2049$ lattice points). Figs.~\ref{cubic-phase}(a) and \ref{cubic-phase}(b) depict the phase diagrams of the IS $(v_b)$ and the Lyapunov exponent $(\lambda_0)$, respectively for the cubic map~\ref{eq:ccml:01}. In Fig~\ref{cubic-phase}, the white lines (contours) indicate  the parameters at which $\lambda_0 = 0$. The regions with $v_b \approx 0.2$ exhibit the presence of SNA in the time series. The white arrow points at the location $(0.6, 1.7543)$ in Fig.~\ref{cubic-phase}(a) for which the SNA nature has been studied in detail. %

\subsection{Characterisation of SNA in the CML of quasiperiodically forced cubic maps}

We also want to confirm that the spatial spread represented in Fig.~\ref{sna-cubic}(b) comes out from the SNA nature of the cubic map~\ref{eq:ccml:01}. To validate this, we evaluate (i) the distribution of FTLEs, (ii) the phase sensitivity, (iii) the partial Fourier sums and (iv) the $0-1$ Test for the CML of quasiperiodically forced cubic maps \ref{eq:cml:01} and \ref{eq:ccml:01} as done before.
\begin{figure}[!ht]
\begin{center}
\includegraphics[width=\linewidth]{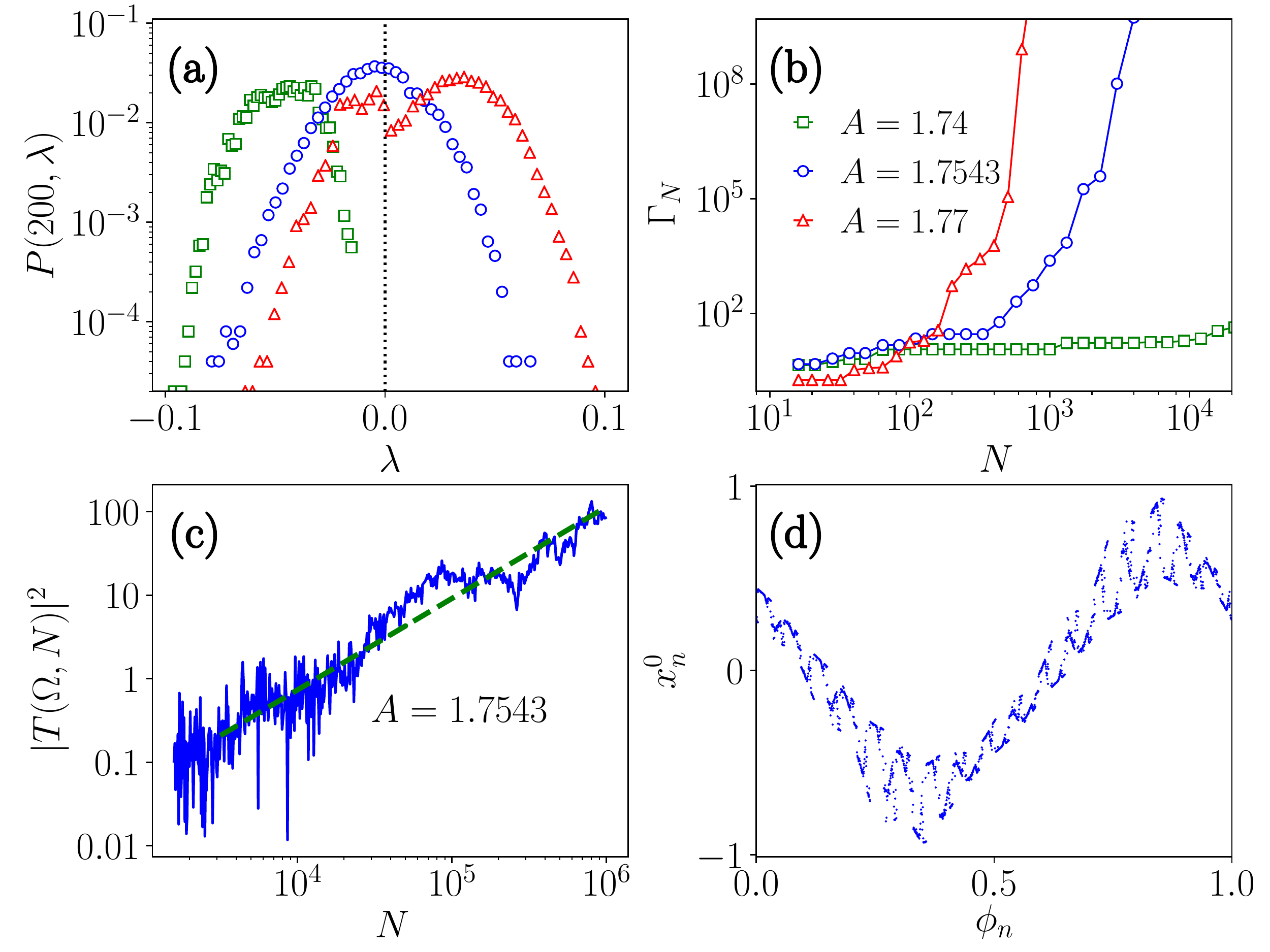}
\end{center}
\caption{
Plots of the characteristic measures of SNA for the CML of forced cubic maps: (a) the distribution of finite time Lyapunov exponents (FTLE) $P(200,\lambda)$, (b) phase sensitivity function $\Gamma_N$ versus $N$, (c) $\vert T(\Omega, N)\vert^2$ showing singular continuous spectrum, and (d) projection of the SNA in the $(x^{0} - \phi)$ plane for  $F = 0.6$ and $A = 1.7543$.}
\label{cubic-all}
\end{figure} %
We analyse the distribution of the largest FTLE for the parameters used in Fig.~\ref{sna-cubic}. In Fig.~\ref{cubic-all}(a), we  show the distribution,  $P(200, \lambda)$ of FTLE as computed using standard procedure~\cite{Wolf1985} for the three choices of $A$ along with $Q$ and $F$ as indicated in Fig.~\ref{sna-cubic}. The largest FTLE is distributed around $-0.05$ as indicated by green squares for $A = 1.74$, while it is distributed around a small negative value $(\approx -0.001)$ as marked by blue-circles  for $ 1.7543$. For $A=1.77$, it is distributed over dominant positive values (red-triangles). The outcome confirms that the isolated map \ref{eq:ccml:01} exhibits SNA  for the choice $A = 1.7453$.

Next, we examine the phase sensitivity of the time series from the estimates of $\Gamma_N$. The variation of $\Gamma_N$ with respect to $N$ follows a similar pattern as seen in the case of CML of forced logistic maps \ref{eq:cml:02} earlier. In Fig.~\ref{cubic-all}(b) we plot $\Gamma_N$ as a function of $N$ for the three sets of parameters considered above, namely $A \in \{1.74, 1.7543, 1.77\}$ with $F = 0.6$. The phase sensitivity function $\Gamma_N$ obeys a power-law relation, i.e. $N^\mu$, in the case of SNA, while it grows exponentially with $N$ for the chaotic case. In the regular (nonchaotic) regime, for example with $A = 1.74$,  $\Gamma_N$ is bounded.  %
\begin{figure}[!ht]
\begin{center}
\includegraphics[width=0.99\linewidth]{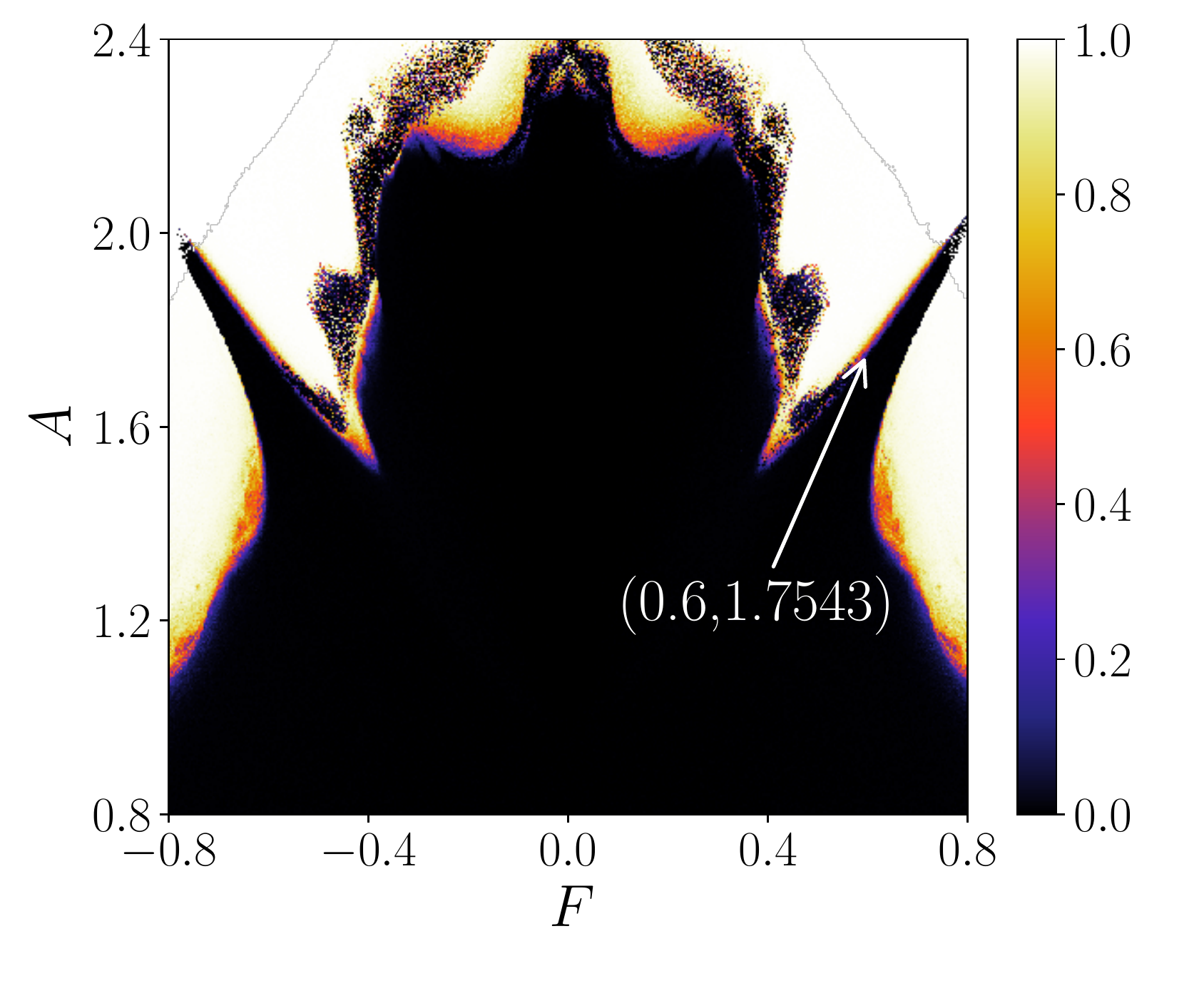}
\end{center}
\caption{Phase diagram of the CML with quasiperiodically forced cubic maps \ref{eq:cml:01} and \ref{eq:ccml:01} in the $F-A$ plane showing density contours (heat map) of $\mathcal K$ \ref{eq:z1:03}. SNA is identified in the regions of $\mathcal K \in (0.2,0.8)$. The white arrow locates the pair of parameters $(F, A)$ used to  demonstrate the SNA in the CML.} 
\label{fig-z1test-cubic}
\end{figure}%

Proceeding further, we  sudy the discrete Fourier transform of the time series $x_n^j$ by estimating the partial Fourier sums from Eq.~\ref{eq:fsum}. In Fig.~\ref{cubic-all}(c), we show the spectrum of partial Fourier sums for $A=1.7543$ for which it displays singular continuous spectrum, with exponent $\beta \approx 1.1 > 1$. This singular continuous spectrum  confirms  the presence of SNA. Fig.~\ref{cubic-all}(d) delineates the projection of the strange non-chaotic attractor in the $x^0 - \phi$ plane for $A= 1.7543$. We note that this projection closely resembles the one created through the Heagy-Hammel route in the case of isolated map~\cite{Venkatesan2001}.

We also carry out the $0-1$ test from the time series $\{x_i^j,\, \, i = 1, 2, \ldots, N\}$ with $j=0$ and length $N=20000$. The $\mathcal K$ values  given in \ref{eq:z1:02} are calculated as $0.002$, $0.4$ and $0.94$ for $A = 1.74$, $1.7543$ and $1.77$, respectively with $Q = 0$, $F = 0.6$ and $\kappa = 0.5$. The $0-1$ test also confirms the presence of SNA for $A = 1.7543$. Further, to provide a broad picture on the dynamics, we plot a phase diagram in the $F-A$ plane by applying the $0-1$ test. Fig.~\ref{fig-z1test-cubic} shows the $0-1$ measure $\mathcal K$ computed from the time series of the CML of cubic maps \ref{eq:cml:01} and \ref{eq:ccml:01} for $F \in (-0.8, 0.8)$ and $A \in (0.8, 2.4)$ in a grid of size $321 \times 321$. For this purpose, we consider the time series $\{x_i^j,\, \, i = 1, 2, \ldots, N\}$ with $j=0$ and length $N=10000$. The bright spots in Fig.~\ref{fig-z1test-cubic} indicate the regions in the parameter space where SNA can be found. This outcome also agrees very well with the two phase diagrams given in Fig.~\ref{cubic-phase} that is obtained from the OTOC.

In the above, we have extended our study on the spread of the OTOC in the CML of quasiperiodically forced cubic maps. We identified that in the CML with forced cubic maps, the OTOC follows a similar dynamics as that of logistic maps discussed in Sec.~\ref{sec:cml:1}. In the nonchaotic case, the initial spread ceases after a finite number of iterations, and it shows ballistic spread with light-cone like structure in the chaotic regime. Again the CML shows a distinct on and off kind of spread in the SNA regime. The SNA nature of the CML is further confirmed from the analysis of the distribution of FTLEs, phase sensitivity, partial Fourier sums, and $0-1$ test. These results are supplemented by the two-parameter phase diagrams for the identification of different dynamical regimes based on the IS, FTLE, and the $0-1$ test.

\section{Summary}
\label{sec:summary}

We studied the spatial spread of the out-of-time-ordered correlators (OTOC) in coupled map lattices of quasiperiodically forced maps that exhibit strange nonchaotic attractors. In particular, we investigated the role of strange non-chaotic attractor on the spatial spread of the OTOC with the aid of certain characteristic measures, namely, the instantaneous speed  (IS) of the spread and the finite-time Lyapunov exponents (FTLEs). We also provided a wider spectrum of the various dynamical regimes in two-parameter phase diagrams by computing the IS and FTLEs. The SNA properties of the isolated maps are reflected in the CML. Interestingly, the OTOC has a characteristic {\it on and off} spread for SNA. This characteristic spread of the OTOC can be seen as a manifestation of the SNA in spatially extended systems.  Further, we carefully examined the presence of SNA in the CML using established measures such as distribution of finite-time Lyapunov exponents, phase sensitivity, partial Fourier sums, and $0-1$ test. Further, we noticed an interesting feature that the dependence of FTLEs on the IS for the quasiperiodically forced maps exhibiting SNAs follows an identical form, which differs from earlier studies.

\acknowledgments

The work of P.M. is supported by CSIR under Grant No. 03(1422)/18/EMR-II, DST-SERB under Grant No. CRG/2019/004059, DST-FIST under Grant No. SR/FST/PSI-204/2015(C), MHRD RUSA 2.0 (Physical Sciences) and DST-PURSE (Phase-II) Programmes. 
The work of M.S. forms a part of a research project sponsored by CSIR, India under Grant No. 03(1397)/17/EMR-II. M.S. also acknowledges MHRD RUSA 2.0 (Physical Sciences) and DST-PURSE (Phase-II) Programmes for providing financial support.

%
\end{document}